\documentclass{aa}
\usepackage{graphicx}
\usepackage{txfonts}
\usepackage{hyperref}
\begin{document}

\title{Fluorescent Fe K line emission of $\gamma$~Cas stars}
\subtitle{II. Predictions for magnetic interactions and white-dwarf accretion scenarios}
\author{G.\ Rauw\inst{1}} 
\offprints{G.\ Rauw}
\mail{g.rauw@uliege.be}

\institute{Space sciences, Technologies and Astrophysics Research (STAR) Institute, Universit\'e de Li\`ege, All\'ee du 6 Ao\^ut, 19c, B\^at B5c, 4000 Li\`ege, Belgium}
\date{Received date/Accepted date}

\abstract
    {About 12\% of the early-type Be stars, which are not known X-ray binaries, exhibit an unusually hard and bright thermal X-ray emission. The X-ray emission of these so-called $\gamma$\,Cas stars could result from accretion onto a white dwarf companion or from magnetic interactions between the Be star and its decretion disc.}
    {Exploring the full power of high-resolution X-ray spectroscopy of $\gamma$~Cas stars requires the comparison of observations of the fluorescent Fe K$\alpha$ emission lines near $\sim 6.4$\,keV with synthetic lines simulated for both scenarios.}
    {We computed synthetic profiles of this line complex within the framework of the magnetic interaction and the accreting white dwarf scenarios. For the latter, we further distinguished between accretion onto a non-magnetic and a magnetic white dwarf. The various models account for different reservoirs of reprocessing material: the Be circumstellar decretion disc, the Be photosphere, an accretion disc around the putative white dwarf companion, a magnetically channelled accretion flow, and the white dwarf photosphere.}
    {We find considerably different line properties for the different scenarios. For a non-magnetic accreting white dwarf, the global Fe K$\alpha$ complex is extremely broad, reaching a full width of 140\,eV, whilst it is $\sim 40$\,eV for the magnetic star-disc interaction and the magnetic accreting white dwarf cases. In the magnetic star-disc interaction, the line centroid is expected to follow the orbital motion of the Be star, whereas it should move along with the white dwarf in the case of an accreting white dwarf. For $\gamma$~Cas, given the $\sim 15 \times$ larger amplitude of the white dwarf orbital motion, the shift in position for an accreting white dwarf should be easily detectable with high-resolution spectrographs such as {\it Resolve} on {\it XRISM}, but remains essentially undetectable for the magnetic star-disc interaction.}
    {Upcoming high-resolution spectroscopy of the fluorescent Fe K$\alpha$ emission lines in the X-ray spectra of $\gamma$~Cas stars will offer important insights into the properties of the primary X-ray source and of the illuminated material, allowing us to distinguish between the competing scenarios.}

\authorrunning{G.\ Rauw et al.}
\titlerunning{Fe K$\alpha$ line in $\gamma$~Cas stars}
\maketitle
\section{Introduction}
The star $\gamma$~Cassiopeiae (HD\,5394, B0.5\,IVe) is well known as the prototype of classical Be stars, which are main-sequence or giant B-type stars that display or have displayed H\,{\sc i} Balmer emission lines in their spectra \citep{Riv13}. Half a century ago, the {\it Small Astronomy Satellite} ({\it SAS-3}) discovered a hard and moderately bright X-ray emission associated with $\gamma$\,Cas \citep{Jer76,Mas76}. The properties of this X-ray emission, which is much brighter and harder than in the majority of  Be stars but significantly fainter than in Be high-mass X-ray binaries (BeHMXBs), are a puzzle in stellar X-ray astrophysics \citep{Smi16,Naz25}. For a long time, $\gamma$\,Cas was the only object known to display these properties. However, over the past two decades, more than 30 Be stars with similar X-ray properties have been discovered \citep[e.g.][]{Smi06,Lop06,Naz18,Naz20,Naz22,Naz23}. Rather than an oddball, $\gamma$~Cas is the prototype of a new class of objects, accounting for $\sim 12$\% of the early-type Be stars \citep{Naz23}. The defining characteristics of these $\gamma$\,Cas stars are X-ray luminosities (in the 0.5 -- 10\,keV band) ranging from $L_{\rm X} \sim 4 \times 10^{31}$ to $1.6 \times 10^{33}$\,erg\,s$^{-1}$, $\log{(L_{\rm X}/L_{\rm bol})}$ between $\sim -6.2$ and $-4$, and an X-ray emission arising in a thermal plasma with $kT \geq 5$\,keV \citep{Naz23}.

Observationally, $\gamma$~Cas and a number of its comrades were found to be single-lined spectroscopic binary systems with orbital periods between weeks and months \citep[][and references therein]{Naz22a}. Existing orbital solutions allow for the unseen companions to be stripped helium subdwarfs\footnote{\citet{Lan20} proposed a collision between the wind of a hot stripped star and the Be disc as the origin of the hard X-ray emission. However, this scenario was found to be in contradiction with observations \citep{Naz22}.}, white dwarfs (WDs), or neutron stars (NSs). The role of the compact companion in the generation of the peculiar X-ray emission remains currently debated. Indeed, the scenarios proposed to explain the X-ray emission can be divided into two broad categories: accretion onto a compact companion or magnetic star-disc interactions.

In the first category, 'classical' accretion onto a NS \citep{Whi82} as in BeHMXBs was excluded because of the thermal nature of the emission, its comparatively low luminosity, and the lack of X-ray pulsations and bursts \citep[see e.g.][and references therein]{Smi16}. \citet{Pos17} proposed that the companion could be a rapidly spinning magnetic NS in the propeller regime with the rapid rotation of the magnetosphere prohibiting accretion. However, this scenario was refuted based on evolutionary as well as observational considerations \citep{Smi17,Rau24}. As an alternative to a NS, accretion onto a WD companion was put forth \citep[e.g.][and references therein]{Mur86,Ham16,Tsu18,Tsu23,Gie23,Toa25}. If this scenario were confirmed, $\gamma$\,Cas stars could populate the category of Be + WD systems predicted by binary evolution population synthesis models but which are currently largely under-represented in the observations \citep{Sha14}.

The second category attributes the hard X-ray emission to interactions between the Be star and its decretion disc, without any direct intervention of the companion star. Motivated by observed correlations between X-ray and UV variability, which indicate that the two radiations arise close to each other, \citet{Smi98} and \citet{Smi99} elaborated the star-disc magnetic interaction scheme \citep[see also][]{Rob00,Mot15,Smi16,Smi19}. This scenario attributes the hard X-ray emission to magnetic reconnections between small-scale magnetic fields at the Be star surface, such as predicted by \citet{Can11}, and a magnetic field rooted in the Be disc.

Whilst $\gamma$\,Cas stars have been observed with nearly every past or existing X-ray telescope, high-resolution X-ray spectroscopy of these objects, especially at energies between 6 and 7\,keV, is still in its infancy. This energy range is especially relevant as it contains several emission lines with complementary diagnostic potentials. The He-like Fe\,{\sc xxv} lines near 6.7\,keV and the H-like Fe\,{\sc xxvi} Ly$\alpha$ line at 6.97\,keV provide information about the very hot plasma that makes up the primary source of hard X-rays. The fluorescent Fe K$\alpha$ line near 6.4\,keV in turn probes the properties of both the primary hard X-ray source and the illuminated cool fluorescent material. Previous X-ray telescopes lacked the spectral resolution to resolve the details of those lines. Thanks to the {\it Resolve} X-ray microcalorimeters on board the {\it X-ray and Imaging Spectroscopy Mission} \citep[{\it XRISM},][]{Ish22}, this situation is changing as the first high-resolution spectra of $\gamma$\,Cas stars in this energy range are becoming available. Further progress will come with the {\it X-ray Integral Field Unit} \citep[{\it X-IFU},][]{Pei25} on the European Space Agenecy's future {\it Athena} X-ray observatory.

Therefore, it is timely to investigate the diagnostic potential of the Fe K$\alpha$ line in $\gamma$\,Cas stars. As a first step, \citet[][hereafter Paper I]{Rau24} modelled the fluorescent line expected in the \citet{Pos17} propeller NS scenario. This study showed that the theoretical predictions could not be reconciled with the observed line strengths and absorbing column densities. In the present paper, we focus on the magnetic star-disc interaction and the accreting WD scenarios. Section\,\ref{sect2} recalls the fundamental concepts of fluorescence, while Sect.\,\ref{Bedisc} describes the circumstellar environment of the Be star and its putative WD companion. Fluorescent Fe K$\alpha$ lines from the star-disc interaction scenario are discussed in Sect.\,\ref{stardisc}, while Sect.\,\ref{accWD} provides the results for accreting WDs. Finally, Sect.\,\ref{conclusion} presents our conclusions.  

\section{Modelling fluorescent emission \label{sect2}}
When a volume of relatively cool gas is illuminated by a bright and hard X-ray source, some atoms and ions in the gas undergo K-shell photoionisation by photons with energies exceeding that of the K-shell ionisation edge. Provided there remain electrons on the L or M shell, the ensuing vacancy on the K shell is filled by one of these electrons. The excess energy of this electron is given away either via Auger autoionisation or by the emission of a fluorescent line photon. The probability that the process leads to fluorescent line emission instead of Auger autoionisation is called the fluorescent yield \citep[e.g.][]{Pal03}. The fact that Fe atoms and ions have relatively high fluorescent yields of about 0.3 \citep{Kal04} makes the Fe K$\alpha$ fluorescence line a rather prominent feature.

In $\gamma$~Cas stars, several media provide reservoirs of cool material that can contribute to the fluorescent line emission. In the framework of the scenarios that we study here, these are the Be decretion disc, the Be photosphere, the circumstellar environment of the WD, and the WD atmosphere.

\subsection{Fluorescence by circumstellar matter}
Our approach to model the fluorescent emission from the circumstellar material closely follows the method outlined in Paper I. The primary X-ray spectrum is described by an APEC optically thin thermal plasma model \citep{Smi01} extended  by a bremsstrahlung continuum at energies above 100\,keV. The energies of Fe K$\alpha$ and K$\beta$ lines as well as the branching ratios between those lines are taken from \citet{Yam14} whilst fluorescent yields are from \citet{Kal04}.

We briefly recall the main concepts of our method adopting the formalism for a point-like source of primary X-ray photons. Further details can be found in Paper I. The Fe K$\alpha$ emissivity of a cell of material, of volume $dV$ and hydrogen particle density $n_{\rm H}(\vec{r})$, located at a position vector $\vec{r}$ from the centre of the Be star, and at a position $\vec{d}$ from the illuminating primary X-ray source can be written
\begin{eqnarray}
  j_{K\alpha}(\vec{r},\vec{d}) & = & E_{K\alpha}\,z_{\rm Fe}\,x_{\rm ion}\,\omega_{\rm ion+1}\,n_{\rm H}(\vec{r}) \nonumber \\
  & & \int_{\rm E_{thres}}^{\infty} \sigma_{\rm K-shell}(E)\,f(E,\vec{d})\,\exp{(-\tau(E,\vec{d}))}\,dE  \label{jline}
  ,\end{eqnarray}
where $\sigma_{\rm K-shell}$ and $E_{\rm thres}$ are respectively the K-shell photo-ionisation cross section and the ionisation threshold energy, both evaluated according to \citet{Ver95}, $E_{K\alpha}$ is the energy of the K$\alpha$ line \citep{Yam14}, $z_{\rm Fe}$ is the (solar) abundance of iron by number with respect to hydrogen \citep{Asp09}, $x_{\rm ion}$ is the relative abundance of a specific Fe ion, $\omega_{\rm ion+1}$ is the K$\alpha$ fluorescent yield \citep{Kal04}, $f(E,\vec{d})$ is the photon flux per unit energy at position $\vec{d}$, and $\tau(E,\vec{d})$ is the optical depth along the line joining the cell to the primary X-ray source.

For an observer located in the direction $\vec{n}$, the contribution of the cell of volume $dV$ to the total line luminosity is expressed as
\begin{equation}
  dL_{K\alpha}(\vec{r},\vec{d}) = j_{K\alpha}(\vec{r},\vec{d})\,\exp{(-\tau(E_{K\alpha},\vec{r},\vec{n}))}\,dV\,{\rm obs}(\vec{r},\vec{n})
,\end{equation}
where ${\rm obs}(\vec{r},\vec{n})$ is equal to 1 or 0 depending on whether or not the cell can be seen by the observer, and $\tau(E_{K\alpha},\vec{r},\vec{n})$ yields the optical depth at the energy of the K$\alpha$ line integrated from the position $\vec{r}$ along the direction $\vec{n}$.

\subsection{Fluorescence in the stellar atmospheres \label{SSatmo}}
Some part of the hard X-rays illuminate the photosphere of either the Be star or the WD companion and could thus give rise to fluorescent line emission. Expressing the position inside the stellar atmosphere via the Rosseland optical depth, $\tau_{\rm Ross}$, the photosphere can be considered a semi-infinite medium illuminated from above by the primary X-ray source under an angle $\theta_{\rm in}$.

To represent the density structure inside the photospheres, we used the non local thermodynamic equilibrium (nLTE) plane-parallel model atmosphere code TLUSTY \citep[][and references therein]{Hub17a,Hub17b,Hub17c,Hub21}. For the Be star, we used TLUSTY models from the BSTAR2006 grid of \citet{Lan07}. For the WD, we computed dedicated models with version 208 of TLUSTY \citep{Hub21}. The optical depth for incoming X-rays of energy $E$ is given by
\begin{equation}
  \tau_{\rm X}(E,z) = \sigma(E)\,\int^0_{z} n_{\rm H}\,dz'
,\end{equation}
where $\sigma(E)$ is the cross-section of the stellar material per H atom, and $z$ stands for the geometrical position measured from the top of the photosphere ($z < 0$ inside the atmosphere). Using the TLUSTY tables, $z$ can be converted into $\tau_{\rm Ross}$.

Here, we assume that we can treat the illuminating X-rays as a small perturbation that does not strongly alter the density, temperature and ionisation stratifications of the photosphere given by the TLUSTY models. To treat the radiative transfer, we adopt the Feautrier formalism of \citet{Gar10}. The mathematical details are outlined in Appendix\,\ref{Feautrier}. The Feautrier equations are solved for a grid of values of $\mu_{\rm in} = \cos{\theta_{\rm in}}$ from 0.025 to 0.975. The stellar surface is discretised into $360 \times 180$ small cells. Depending on the geometry (point-like or extended) and location of the illuminating source of hard X-rays, and depending on the location of the observer, we evaluate for each of the cells the visibility of the source, the associated value of $\mu_{\rm in}$, and the value of $\mu = \cos{\theta}$, where $\theta$ is the angle between the direction $\vec{n}$ towards the observer and the vertical $z$-axis. Moreover, the rotational velocity of each cell as seen by the observer is used to Doppler-shift the emerging fluorescent photon. All photons received by the observer are then grouped into a line profile according to their energy as seen by the observer.

\subsection{Global Fe K$\alpha$ profiles}
As in Paper I, we first compute synthetic line profiles of individual transitions taking into account the Doppler shift due to the velocity component of the emitting cell along the $\vec{n}$ direction. The K$\alpha$ line consists of an unresolved array of transitions arising from the various combinations of L and K shell electronic configurations \citep{Pal03}. Each individual transition has its specific energy and intrinsic strength. The synthetic profiles of the full Fe K$\alpha$ complex are then obtained by convolving the energy distribution of the transitions with the individual line profile weighted by the intrinsic strength of the transitions.

\section{The circumstellar environment \label{Bedisc}}
To model the fluorescence from $\gamma$\,Cas, we need several parameters of the system. These include the spectral energy distribution of the primary X-ray emission, the orbital parameters of the binary system, the mass and radius of the Be star, and a description of the density structure of the circumstellar material. In Paper I, a set of relevant parameters was compiled. These default parameters are given in Table\,\ref{tab_param}. Here, we added also our default parameters for the WD and its circumstellar environment. 

\begin{table}
  \small
  \caption{Model parameters of $\gamma$~Cas. \label{tab_param}}
  \begin{center}
  \begin{tabular}{l c | l c}
    \hline
    \multicolumn{2}{c|}{Primary X-ray source} & \multicolumn{2}{c}{WD companion}\\
    \hline
    $kT$ (keV) & 12.5 & $M_{\rm WD}$ (M$_{\odot}$) & 0.98\\
    $f_{\rm X}$ (erg\,cm$^{-2}$\,s$^{-1}$) & $1.5 \times 10^{-10}$ & $R_{\rm WD}$ (R$_{\odot}$) & 0.01\\
    \cline{1-2}
    \multicolumn{2}{c|}{Be star} & $T_{\rm eff}$ (kK) & 60 \\
    \cline{1-2}
    $M_*$ (M$_{\odot}$) & 16 & $\log{g}$ & 8.5 \\
    $R_*$ (R$_{\odot}$) & 10.0 & $v_{\rm rot}^{\rm WD}$ (km\,s$^{-1}$) & 25 \\
    \cline{3-4}
    $T_{\rm eff}$ (kK) & 25 & \multicolumn{2}{c}{WD accretion disc}\\
    \hline
    \multicolumn{2}{c|}{Be disc} & $\alpha_{\rm WD}$ & 2.4 \\
    \cline{1-2}
    $i$ ($^{\circ}$) & 45 & $n_0^{\rm WD}$ ($10^{16}$\,cm$^{-3}$) & 3.0 \\
    $n_0$ ($10^{13}$\,cm$^{-3}$) & 2.2 & $\dot{M}_{\rm WD}$ (M$_{\odot}$\,yr$^{-1}$) & $2 \times 10^{-11}$ \\
    \cline{3-4}
    $\alpha$ & 2.5 & \multicolumn{2}{c}{Roche lobe radii}\\
    \cline{3-4}
    $R_{\rm out}$ (R$_{\odot}$) & 176 & Be star & $0.61\,a$\\
    \cline{1-2} 
    \multicolumn{2}{c|}{Binary orbit} & WD & $0.18\,a$  \\
    \cline{1-2}
    $P_{\rm orb}$ (d) & 203.523 \\
    $a$ (R$_{\odot}$) & 375 \\
    \hline 
  \end{tabular}
  \end{center}
  \tablefoot{To convert the X-ray fluxes into luminosities, we adopted a distance of $\gamma$\,Cas equal to 190\,pc \citep{Naz18}. The WD mass, radius, and accretion disc properties are adopted from \citet{Ras25}.}
\end{table}

\begin{figure}
  \resizebox{8.5cm}{!}{\includegraphics[angle=0]{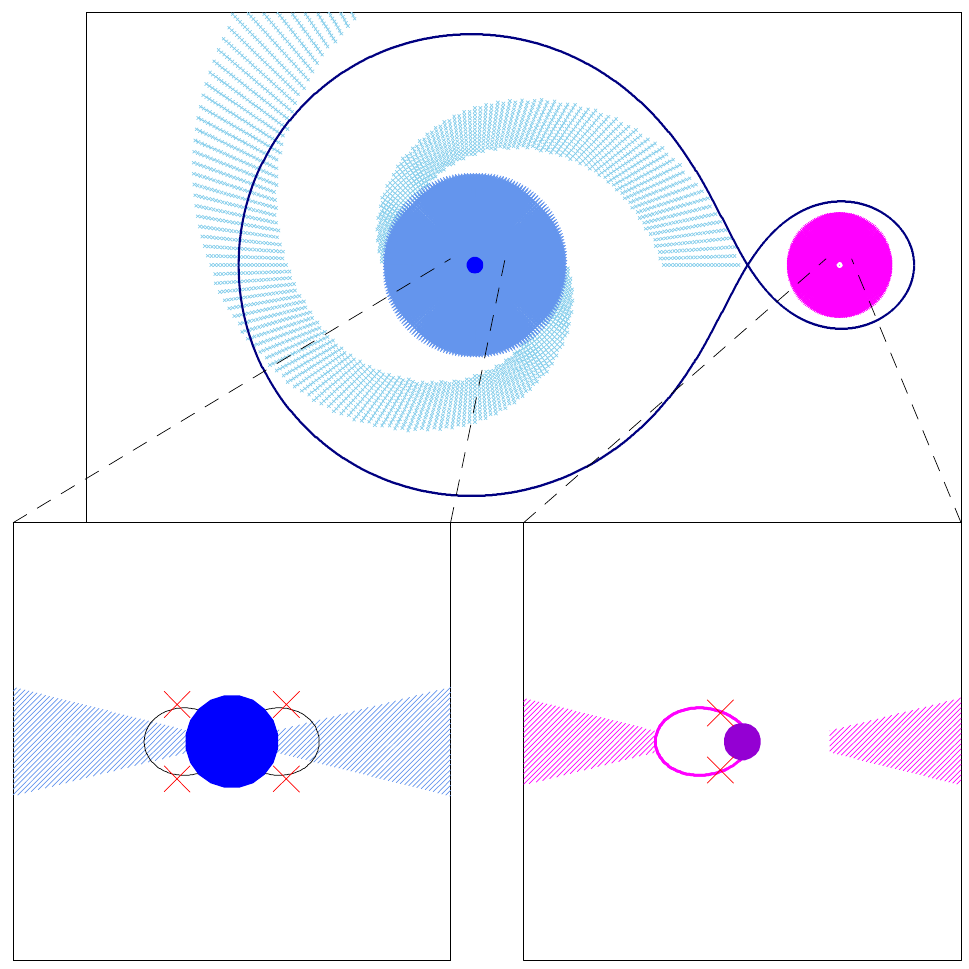}}
  \caption{Upper box: Schematic pole-on view of the Be star (blue dot), its decretion disc and spiral extensions (sky blue), the WD companion, and its truncated accretion disc (magenta). The dark blue contour shows the Roche lobe of the system. The boxes at the bottom provide zooms on vertical cuts through the inner part of the Be decretion disc (left) and the surroundings of the WD (shown by the violet dot) with its magnetically truncated accretion disc (right). The red crosses correspond to the emitting regions of the primary X-rays. The left box illustrates the case of the magnetic star-disc interaction with magnetic field lines that connect the star and the disc. The right box corresponds to an accreting magnetic WD scenario where a magnetically channelled flow connects the truncated accretion disc to the magnetic poles of the WD. \label{sketch}}
\end{figure}

The decretion discs of Be stars in low-eccentricity binary systems are usually expected to be truncated at the 3:1 resonance radius via gravitational interaction with the companion \citep{Oka01,Oka02,Pan16}. This concept has been revised in recent years, giving way to a more complex view of Be binary systems. Indeed, based on smoothed particle hydrodynamics (SPH) simulations, \citet{Mar24} showed that in BeHMXBs the efficiency of tidal truncation depends on the disc aspect ratio. For $H/R \leq 0.1$ tidal truncation was found to operate efficiently, but for larger $H/R$ values its effect is much weaker. Since Be discs are flared, their aspect ratio increases with radius, which implies that disc truncation should be less efficient for longer period binaries which have wider orbital separations. \citet{Mar24} further showed that material that escapes the Roche lobe of the Be star flows along tidal streams towards the compact companion's gravitational sphere of influence where it forms an accretion disc. Similar conclusions were reached by \citet{Rub25} and \citet{Ras25} who showed that instead of a sharp truncation, the density of the decretion disc undergoes a strong, azimuthally dependent decrease over a range of several stellar radii. \citet{Rub25} distinguished five regions in the Be binary system. With increasing distance from the Be star, these are (1) the inner Be disc, which is least perturbed by the companion, (2) a spiral-dominated disc featuring two-armed density waves, (3) a bridge formed by the denser arm that extends into the companion's Roche lobe, (4) a disc-like accretion structure around the companion, and (5) a circumbinary envelope fed by material lost from the system. Figure\,\ref{sketch} provides a sketch of the circumstellar environment of the Be star. 

We assume that the innermost parts of the Be decretion disc can be described as an axisymmetric structure with a hydrogen number density distribution given by 
\begin{equation}
  n_{\rm H}(r,z) = n_0\,\left(\frac{r}{R_*}\right)^{-\alpha}\,\exp{\left[-\frac{1}{2}\,\left(\frac{z}{H(r)}\right)^2\right]}
  \label{Hummel}
\end{equation}
 \citep[e.g.][]{Hum00}. In this relation, $r$ and $z$ are the radial and vertical components of the cylindrical coordinates, $n_0$ is the hydrogen number density at the inner edge of the disc, and $H(r)$ is the disc scale height. The latter quantity varies with $r$ as
\begin{equation}
  H(r) = \frac{c_s(R_*)\,R_*}{v_{\rm rot}(R_*)}\,\left(\frac{r}{R_*}\right)^{\frac{3-p}{2}}
  \label{scaleheight}
  ,\end{equation}
where $c_s(R_*)$ is the speed of sound at the inner edge of the disc. The exponent $p$, introduced by \citet{Kur14} to describe the radial temperature profile of the disc, was taken to be $p = 0.25$.
In the case of our default simulations for $\gamma$~Cas, the aspect ratio remains below 0.1 all the way out to the 3:1 resonance radius \citep[about 176\,R$_{\odot}$,][]{Rau22}, indicating that truncation might remain effective. The transition to an aspect ratio above 0.1 takes place at $r = 194$\,R$_{\odot}$. 

To account for the more complex disc structure described by \citet{Rub25}, we assumed the inner disc to extend out to 8\,R$_*$ and represented the spiral dominated disc, the bridge region and the circumbinary region by two logarithmic spirals (see Fig.\,\ref{sketch}). To mimic the properties found by \citet{Rub25}, both spirals were assumed to arise over a range in azimuthal angle of $30^{\circ}$ and to have a density of 3 times the nominal density expected at this distance (as given by Eq.\,(\ref{Hummel})). Conversely the inter-arm medium was assumed to have a density 0.6 times that given by Eq.\,(\ref{Hummel}). The spiral arms emanate from the disc at the points facing the secondary and opposite to it. The leading spiral arm, which turns into the bridge that connects with the secondary's Roche lobe, stops at the position of the secondary, whilst the trailing arm feeds the circumbinary region (see Fig.\,\ref{sketch}). To avoid introducing additional parameters, we assume that the companion's orbital plane coincides with the equatorial plane of the Be star, and hence the plane of the decretion disc.

\section{X-ray emission from the magnetic interaction scenario \label{stardisc}}
In the context of the magnetic interaction scenario, the X-ray emission is attributed to magnetic reconnection events between localised magnetic fields at the surface of the Be star and magnetic fields arising inside the Be disc. The hot X-ray-emitting plasma would thus be located near the apex of the magnetic loops \citep{Smi98} in 'canopies' above the Be star and close to the innermost parts of the disc \citep{Smi16}. To represent this situation, we consider two rings of material encircling the Be star, one above the equatorial decretion disc, the second one below the disc. In our default model, the rings are assumed to have a radius of $R_s = 1.5\,R_*$ as measured from the Be star's rotation axis, and to be located at a height $Z_s = 0.5\,R_*$ above or below the disc. In the absence of dedicated magneto-hydrodynamic simulations, these values were motivated by the constraint that, in this model, the primary X-ray emission arises near the Be star surface. The rings are discretised into 180 elements of $2^{\circ}$ azimuthal extension. The total X-ray luminosity is distributed equally over those elements. Each element of the ring is assimilated to a point source, and the total fluorescent emission is obtained as the combination of the illumination by these 180 fake point sources.

This assumption implies that our model represents a time-averaged configuration. Indeed, at any given time, localised magnetic spots arising from the subsurface iron convection zone \citep{Can11} are probably not distributed evenly in stellar longitude. Moreover, in the magnetic interaction scenario, the ubiquitous X-ray shots are interpreted as individual reconnection events taking place randomly, which implies a significant temporal variability of the X-ray emission from a specific element. Considering a typical $\sim 50$\,ks integration time, needed to collect high-resolution X-ray spectra, we expect these rapid variations to average out.  

\begin{figure}
  \resizebox{8.5cm}{!}{\includegraphics[angle=0]{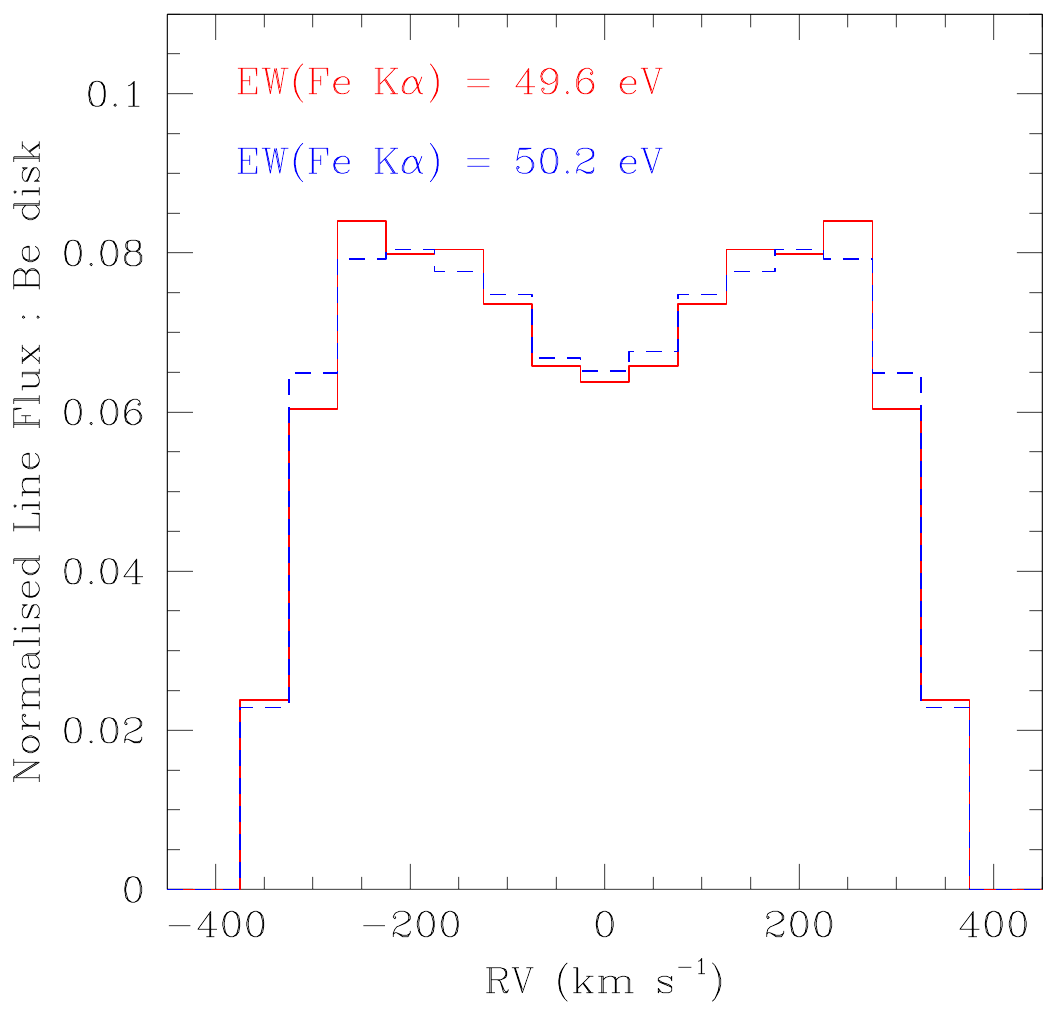}}
  \caption{Normalised individual Fe K$\alpha$ line profile contributed by the Be decretion disc assuming an extended source represented by two rings of hot plasma of radius $1.5\,R_*$ and located $0.5\,R_*$ above or below the Be decretion disc. The line profile is binned into velocity bins of 50 km\,s$^{-1}$ and is normalised by dividing by the total flux over the entire line profile. The spectral energy distribution of the illuminating source is an optically thin thermal plasma model with $kT = 12.5$\,keV. The red histogram corresponds to an axisymmetric disc truncated at $r = 176$\,R$_{\odot}$. The blue histogram yields the results for a disc with spiral extensions as described in the text and computed out to $a = 375$\,R$_{\odot}$. The normalised profiles are displayed in velocity bins of 50\,km\,s$^{-1}$. \label{normprof}}
\end{figure}

As a first ingredient, we consider the fluorescent emission arising from the decretion disc. Because the illumination is maximum over the innermost parts of the disc, we start by assuming that the spiral part of the disc as well as the bridge and other components of the flow play only a minor role. The predicted individual line profile is shown in Fig.\,\ref{normprof}. The individual profile displays a double-peaked morphology reflecting the Keplerian velocity field of the inner disc part. The total equivalent width (EW) of the Be disc fluorescent line\footnote{The fluorescent Fe\,K$\alpha$ line observed in the {\it XMM-Newton} and {\it Chandra} spectra of $\gamma$\,Cas has an EW in the range between 34 and 59\,eV \citep{Rau22}.} is predicted to be 49\,eV whilst the line luminosity amounts to $3.3 \times 10^{30}$\,erg\,s$^{-1}$. Part of the generated Fe K$\alpha$ photons are reabsorbed by the disc material which is mildly optically thick in its inner regions\footnote{Reabsorption is restricted to the inner disc. The contribution of the spiral arms is negligible.}. Indeed, without this re-absorption, the EW would be 70\,eV.  

\begin{figure}
  \resizebox{8.5cm}{!}{\includegraphics[angle=0]{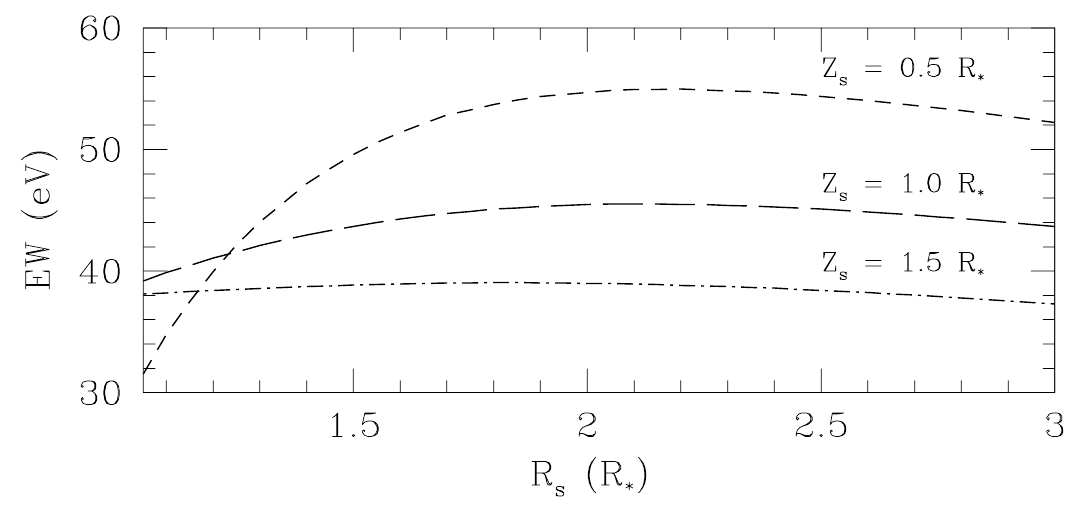}}
  \resizebox{8.5cm}{!}{\includegraphics[angle=0]{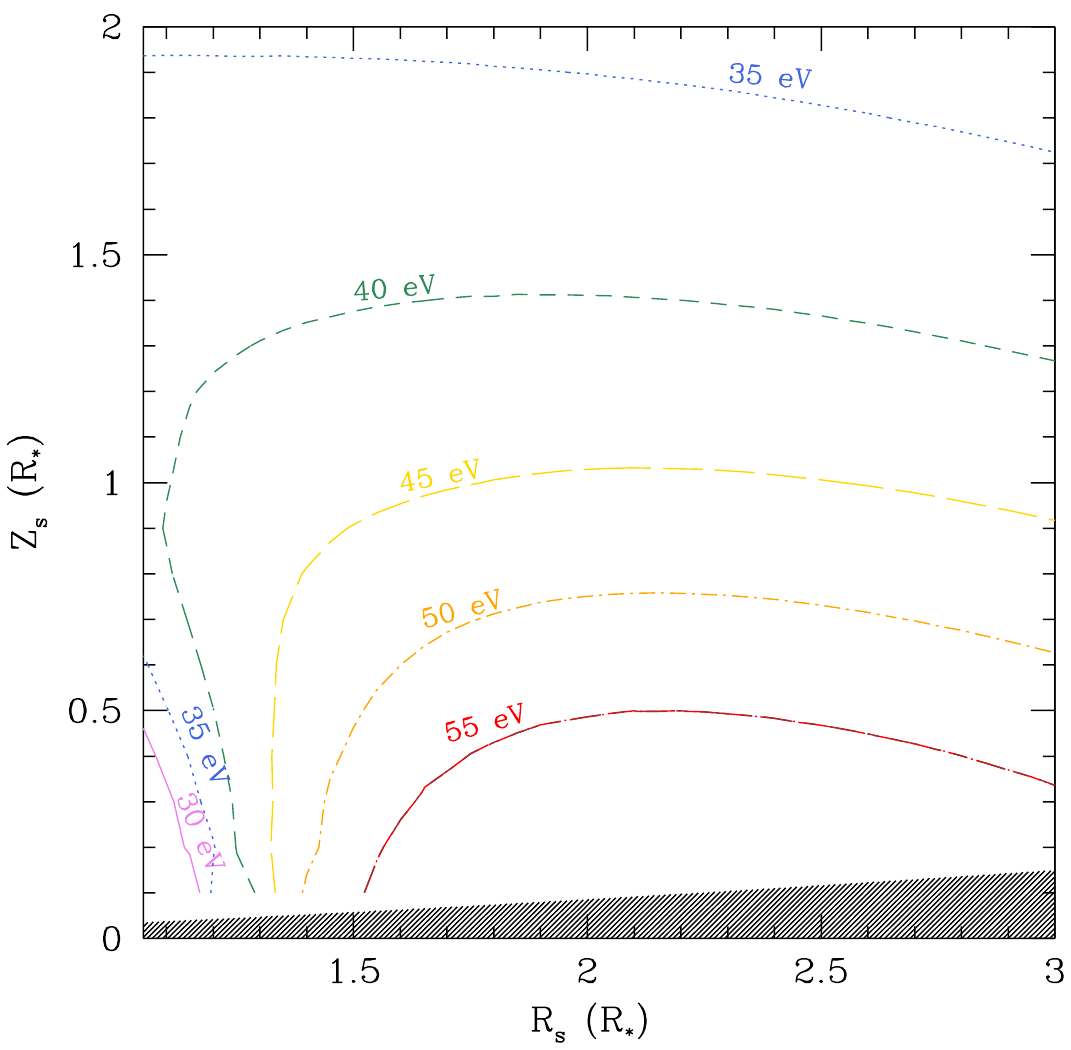}}
  \caption{{\it Top :} EW of the fluorescent Fe K$\alpha$ line contributed by the Be decretion disc in the star-disc interaction scenario as a function of $R_s$ for different values of $Z_s$. {\it Bottom :} EW contours for the same scenario
as a function of $R_s$ and $Z_s$. The primary source is modelled as a uniform axisymmetric ring of radius $R_s$ from the polar axis of the Be star and of altitude $Z_s$ above the equatorial plane of the Be star. The hatched area in the bottom part corresponds to the scale height of the Be decretion disc.\label{EW_grid}}
\end{figure}

In our model of the magnetic interactions, the strength of the fluorescent line depends on the assumed location of the illuminating hot plasma ring. Figure\,\ref{EW_grid} explores the dependence of the EW on the location of the extended primary source. If it is located very close to the Be star and to the decretion disc, it illuminates only a small part of the disc. Moreover this part has the highest density and thus the highest optical depth for absorption of Fe K$\alpha$ photons by the disc material. As a result, we expect the EW of the disc line to be relatively low. Moving to higher altitudes above the disc, that is increasing $Z_s$, the illuminated part of the disc increases, but the ionising flux received by each cell of the disc decreases. Therefore, for $R_s \leq 1.3\,R_*$ and increasing $Z_s$, the EW first increases before it decreases again. Finally, as $R_s$ increases above $1.3\,R_*$, those parts of the disc that are most strongly illuminated by the primary X-ray source are no longer optically thick and fully contribute to the strength of the fluorescence line. Over this part of the parameter space, where optical depth no longer plays a role, the EW is set by two competing trends. On the one hand, the density of the disc material, that most strongly contributes to the fluorescence, decreases with $R_s$, whereas, on the other hand, the extent of the illuminated region increases asymptotically with $R_s$. Hence, for a fixed $Z_s$, EW increases as $R_s$ increases from $1.3\,R_*$ up until $\sim 2\,R_*$ and slowly decreases afterwards as the density of the most-strongly illuminated parts of the disc decreases.

We repeated the calculation of the line due to the Be disc, but this time accounting for the disc structure found by \citet{Rub25} according to our description above. The resulting individual line profile is compared with our simple disc model in Fig.\,\ref{normprof}. The two profiles are nearly identical as are the total EWs. This is because the bulk of the fluorescent emission arises from illumination of the inner part of the disc which is not impacted by the spiral structure. The observer's viewing angle on the spiral arms (and thus their velocity along the observer's sightline) changes as a function of orbital phase, which could lead to variations in the individual line profiles. However, because of their low density and large distance from the primary source, these arms contribute very little to the profile (see Fig.\,\ref{normprof}), and the ensuing variations should be very small.

Beside the Be disc, a second contribution to the fluorescent line comes from the illumination of the Be photosphere. Again the primary X-ray source was discretised into 180 small sectors, each one assimilated to a point source located at a distance $\sqrt{R_s^2 + Z_s^2}$ from the star's centre. We then used the Feautrier method described in Appendix\,\ref{Feautrier} in combination with a TLUSTY model atmosphere of solar metallicity, $T_{\rm eff} = 25\,000$\,K, and $\log{g} = 3.75$, taken from the BSTAR2006 grid \citep{Lan07}, to simulate the photospheric fluorescence.

\begin{figure}
  \resizebox{8.5cm}{!}{\includegraphics[angle=0]{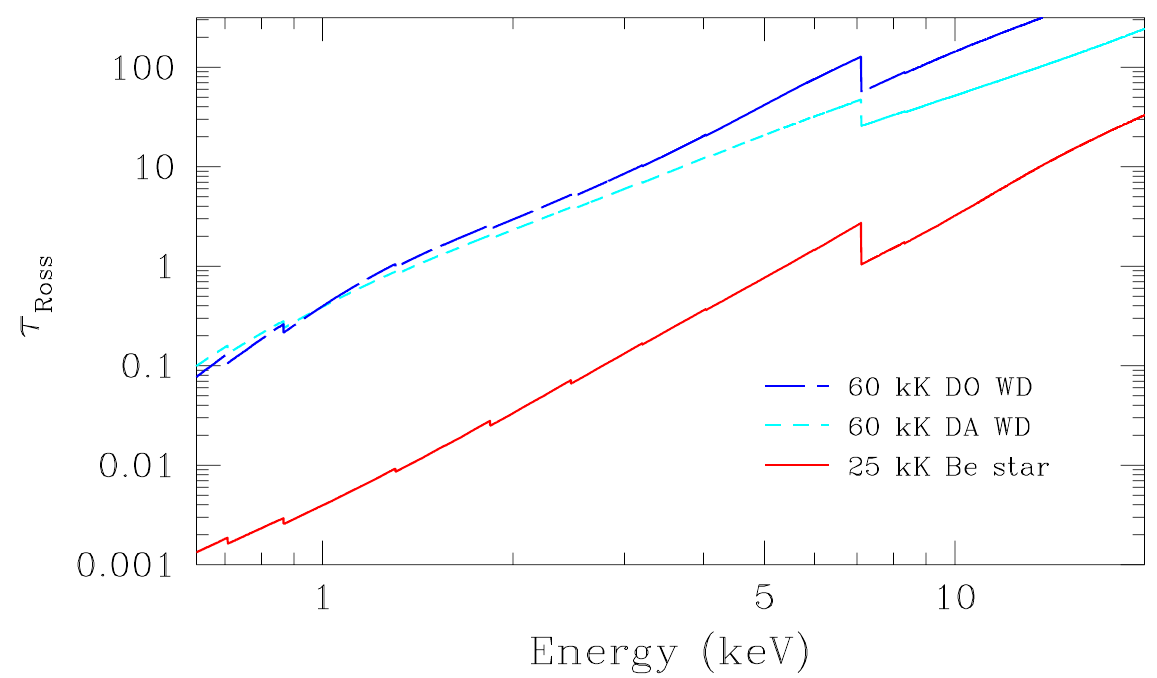}}
  \caption{Rosseland optical depth inside the photosphere where the optical depth for photoelectric absorption of X-rays reaches unity. The continuous red line illustrates the results for the $T_{\rm eff} = 25\,000$\,K, $\log{g} = 3.75$ TLUSTY model adopted for the Be star. The short-dashed cyan line and the long-dashed blue line correspond respectively to DA and DO WD models with $T_{\rm eff} = 60\,000$\,K and $\log{g} = 8.5$.\label{Rosseland}}
\end{figure}

In this model, the bulk of the photoelectric absorption by Fe ions, that is at the 7.1\,keV K-shell ionisation edge, happen to occur at $\tau_{\rm Ross} \simeq 1$ (see Fig.\,\ref{Rosseland}). According to Fig.\,16 of \citet{Lan07}, the dominant ionisation stage of Fe at these depths is Fe\,{\sc iv} ($\geq 90$\,\%) with minor contributions from Fe\,{\sc iii} and Fe\,{\sc v} ($\leq 5$\,\% for each of those species). We can use the ionisation parameter $\xi$, defined as
\begin{equation}
  \xi = \frac{4\,\pi\,F_{\rm X}}{n_e}
  \label{xiEquation}
,\end{equation}
with $n_e$ the electron density, and $F_{\rm X}$ the X-ray flux over the 1 - 1000\,Ry energy domain \citep{Kal04}, to assess the impact of the illumination by a powerful X-ray source on the ionisation balance. This impact is very small for $\log{\xi} < 0$, but strongly increases for $\log{\xi} > 1$ \citep[see Fig.\,6 of][]{Kal04}. At $\tau_{\rm Ross} = 1$, the electron density in the TLUSTY model is $\simeq 4.6 \times 10^{14}$\,cm$^{-3}$. For an extended source, with $\sqrt{R_s^2 + Z_s^2} = 1.58\,R_*$, we obtain $\log{\xi} \leq -5.1$. Therefore, one does not expect significant changes of the ionisation of the photospheric material at the depths where most of the K-shell photoionisation takes place.

\begin{figure}
  \resizebox{8.5cm}{!}{\includegraphics[angle=0]{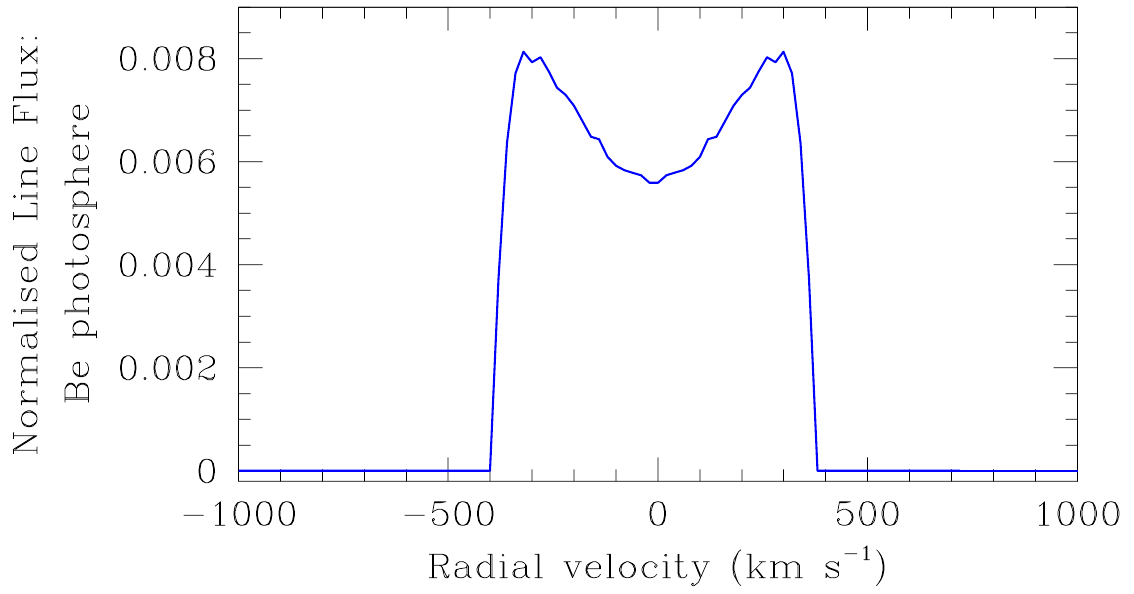}}
  \caption{Individual Fe K$\alpha$ line profile contributed by the Be photosphere. The illuminating source is an extended ring-like source located at $R_s = 1.5\,R_*$ and $Z_s = 0.5\,R_*$. The spectral energy distribution of the illuminating source is an optically thin thermal plasma model with $kT = 12.5$\,keV. \label{profile_photo}}
\end{figure}

The predicted line profile is illustrated in Fig.\,\ref{profile_photo}. The total line luminosity from fluorescence in the Be photosphere amounts to $2.7 \times 10^{27}$\,erg\,s$^{-1}$, and the EW to 0.05\,eV for $R_s = 1.5\,R_*$ and $Z_s = 0.5\,R_*$. This line strength is thus several orders below the disc contribution.\\

\begin{figure}
  \resizebox{8.5cm}{!}{\includegraphics[angle=0]{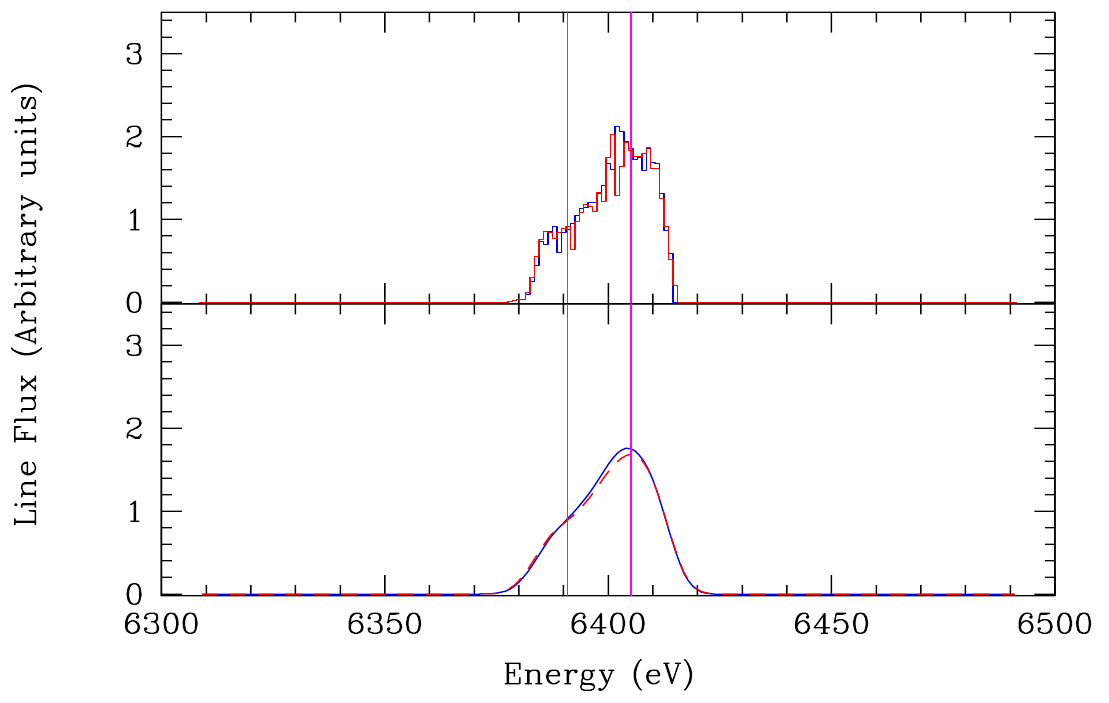}}
  \caption{Top: Global line profile (i.e.\ combining the Be-disc and photospheric contributions and accounting for the full array of transitions) for our default magnetic interaction model, at two opposite quadrature orbital phases. The profiles are shown as histograms binned in energy bins of 1\,eV. Bottom: Convolution of the global line profiles with an instrumental response function of 5\,eV FWHM. The magenta lines yield the theoretical energies of the centroids of the Fe\,{\sc v} K$\alpha$1 and K$\alpha$2 subgroups \citep{Pal03} assuming Fe {\sc iv} target ions. \label{global_ext}}
\end{figure}

Accounting for the fact that the Fe K$\alpha$ emission consists of a complex array of transitions, we convolved this array with the individual profiles obtained after summing the disc line and photospheric profiles. The result is illustrated in Fig.\,\ref{global_ext}. We compare the synthetic profiles expected for the two quadrature orbital phases. The blue (resp.\ red) histogram yields the global Fe K$\alpha$ fluorescent line profile with energy bins of 1\,eV at the quadrature phase when the Be primary moves towards us (resp.\ away from us). We further convolved these synthetic profiles with an instrumental profile whose full width at half maximum (FWHM) would be equal to 5\,eV. The results are shown in the bottom panel of Fig.\,\ref{global_ext}. Owing to the widths of the individual profiles, the resulting blend does not allow one to distinguish the K$\alpha_1$ and K$\alpha_2$ subgroups. K$\alpha_2$ only appears as an extension of the red wing of the blend. Moreover, with a semi-amplitude of 4.1\,km\,s$^{-1}$ \citep[e.g.][]{Rau22}, the radial velocity (RV) motion of the Be star is too small to induce a measurable shift in the overall line profiles, even when considering the most extreme orbital phases. The energy shift between the overall lines at those phases should be only $\sim 0.2$\,eV.

At this stage, we thus conclude that the magnetic interaction scenario can reproduce the observed EWs of the fluorescent Fe K$\alpha$ emission lines in $\gamma$\,Cas stars. An important observational test for this scenario stems from the fact that the line morphology should not vary in a systematic way with orbital phase of the companion and that the line centroid should not display significant RV variations. 

\section{X-ray emission from accreting white dwarfs \label{accWD}}
Accreting WDs are found in two important categories of interacting binaries: cataclysmic variables (CVs) where the WD accretes material via Roche lobe overflow of a main-sequence low-mass star, and symbiotic stars where the mass donor is a red giant. X-ray emission of accreting WDs \citep[for reviews see][]{Muk17,Pag22} is powered by accretion and/or nuclear fusion (either explosively, as in novae, or more steady). Nuclear burning at the surface of a WD manifests itself in the soft X-ray domain (below about 0.5\,keV) as a so-called supersoft source. A few Be + WD systems have been identified in the SMC whilst they were undergoing an outburst \citep{Ken21}. These systems displayed bright ($L_{\rm X} \simeq 10^{36}$\,erg\,s$^{-1}$) supersoft emission attributed to an enhanced accretion activity \citep[rather than nuclear burning on the WD surface,][]{Ken21}. This situation differs significantly from $\gamma$~Cas stars which rather exhibit a moderately bright, hard, and optically thin thermal X-ray emission that is more reminiscent of CVs in quiescence. 

A key parameter of known accreting WDs is the strength of the WD's magnetic field. If it is strong enough \citep[$> 10^5$\,G,][]{Pag22}, the magnetic field takes control of the inflowing material which either entirely prevents the formation of an accretion disc or disrupts the accretion disc at the Alfv\'en radius of the WD, $R_A$. Systems lacking an accretion disc are dubbed polars, whereas those with truncated discs are called intermediate polars (IPs). The former feature WDs with magnetic fields of about $7 \times 10^6$ -- $2 \times 10^8$\,G, whereas the latter have WD field strengths between $10^5$ and $10^7$\,G \citep[e.g.][]{Ram07}. In such systems, the magnetically channelled accretion flow undergoes a hydrodynamic shock at some height, $h_s$, above the WD surface near the magnetic poles. The hard primary X-ray emission arises in shock-heated plasma. The hardness of the corresponding emission depends directly on the WD mass \citep{Muk17}. Frequently, this standoff shock is located close enough to the WD surface, typically at some fraction of the WD radius, $R_{\rm WD}$, so that the solid angle illuminated by the downward radiation is large \citep{Hai16,Tsu18,Lop19}. Assuming a point-like primary X-ray source, this solid angle is given by
\begin{equation}
  \Omega = 2\,\pi\,\left(1 - \sqrt{1 - \frac{R_{\rm WD}^2}{(h_s + R_{\rm WD})^2}}\right).
\end{equation}
For low values of $h_s$, the WD surface intercepts and reprocesses nearly half of the primary X-rays: they are either reflected, Compton scattered or absorbed to ionise iron atoms, resulting in fluorescent line emission. This allows for rather strong Fe K$\alpha$ lines with EWs up to 100 and 200\,eV in polars or IPs. By contrast, quiescent dwarf nova systems with low accretion rates ($\leq 10^{-10}$\,M$_{\odot}$\,yr$^{-1}$) usually display weaker fluorescent lines \citep{Ezu99,Pan05,Byc10,Eze15,Xu16}.

In this section, we consider the two scenarios, non-magnetic and magnetic WD in terms of the ensuing fluorescent Fe K$\alpha$ line. Unfortunately, very little is known about the nature and properties of the companion of $\gamma$~Cas. Assuming it to be a WD, its formation did not proceed along the path followed by classical WDs that are descendants of single low-mass stars. Instead, it would be the outcome of binary evolution, descending from a stripped hot subdwarf. Therefore, its temperature and chemical composition could differ from those of classical WDs. Similar uncertainties apply to the WD's rotational velocity. In polars, one usually considers that the WD rotates synchronously with the orbital motion \citep[e.g.][]{Ram01}. Given the wide orbital separation and relative youth of $\gamma$~Cas, synchronisation, which would imply a WD equatorial rotational velocity of $2.5$\,m\,s$^{-1}$, is unlikely. From the observational viewpoint, several authors attempted to identify X-ray pulsation periodicities in the light curves of $\gamma$~Cas stars that could reflect the rotational period of the putative WD companion. So far, no stable periodicity was found for $\gamma$~Cas itself \citep{Tsu18}, but occasional transient signals have been reported for other $\gamma$~Cas stars. For instance, \citet{Lop06} found an oscillation in the {\it XMM-Newton} light curve of HD~161103 with a period of about 3200\,s, and \citet{Hue24} reported a 3400\,s periodicity in one {\it Chandra} observation of $\pi$~Aqr. If these signals truly reflect the rotation of a WD companion, they would hint at low rotational velocities. In our calculations, we assumed $v_{\rm rot}^{\rm WD} = 25$\,km\,s$^{-1}$, corresponding to a rotation period of 1750\,s. For the WD's mass and radius, we adopted the parameters used by \citet{Ras25}: $R_{\rm WD} = 0.01$\,R$_{\odot}$, $M_{\rm WD} = 0.98$\,M$_{\odot}$. The SPH simulations carried out by \citet{Ras25} predict an accretion rate of the WD in $\gamma$\,Cas of $2 \times 10^{-11}$\,M$_{\odot}$\,yr$^{-1}$, and we used this value as our baseline.

\subsection{Accretion onto a non-magnetic WD}
In non-magnetic CVs, accretion-powered X-rays are emitted in the boundary layer between the Keplerian accretion disc and the WD where the material slows down from a Keplerian rotation velocity near 3000\,km\,s$^{-1}$ to the significantly slower rotational velocity of the WD \citep{Muk17,Pag22}. The hot gas then forms a hot corona with emission of X-rays up to 20\,keV \citep{Pag22}.

For a non-magnetic WD secondary, the accretion disc extends from the radius of the WD's Roche lobe down to the surface of the WD. We assimilated the primary X-ray source, that is the whole boundary layer, to a point-like source. To compute the fluorescence produced by an accretion disc illuminated by a compact object, we modified the code used in Sect.\,\ref{stardisc} by implementing a variable mesh step in the radial direction of the disc. This was necessary to account for the existence of two very different length scales which are $R_{\rm WD}$, and the orbital separation, $a$. In the innermost regions of the WD accretion disc, the radial step was set to 0.05\,$R_{\rm WD}$ and it was increased to 0.1\,$R_{\rm WD}$ (starting at $r = 10\,R_{\rm WD}$), 1\,$R_{\rm WD}$ (starting at $r = 100\,R_{\rm WD}$) and finally to $\frac{r_{\rm out} - 500\,R_{\rm WD}}{500}$ (starting at $r = 500\,R_{\rm WD}$).

From dedicated SPH simulations of the $\gamma$\,Cas system, and assuming a WD nature of the secondary component, \citet{Ras25} showed that the accretion disc around the WD follows a density profile as in Eq.\,(\ref{Hummel}) with $\alpha_{\rm WD} = 2.4$ and with a density $\rho = 10^{-14}$\,g\,cm$^{-3}$ at a distance of 700\,$R_{\rm WD}$ from the WD. If we extrapolate this WD disc density profile inwards to 1\,$R_{\rm WD}$, the hydrogen particle density at the inner boundary of the disc becomes $n_0^{\rm WD} = 3 \times 10^{16}$\,cm$^{-3}$. 

As a first step, we considered that the primary source is located at the geometric centre of the disc. Despite the proximity of the primary source to the reprocessing material and despite the great density at the inner disc boundary, the EW of the fluorescent line generated in the accretion disc amounts to $\leq 0.6$\,eV. There are two reasons for this extremely weak line. On the one hand, the solid angle occupied by the WD accretion disc as seen from the point source is very small (0.074\,sr, that is 0.6\% of the complete $4\,\pi$ solid angle). On the other hand, the density in the inner regions of the disc is so high that it becomes optically thick in the radial direction even at energies of the Fe K-shell ionisation edge. Therefore, most of the ionising X-ray radiation only penetrates over a short radial distance into the disc. Hence, in such a configuration most parts of the disc do not contribute to the formation of the fluorescent line.

It is probably more realistic to assume that the accretion boundary layer and the associated hot corona extend out of the plane of the disc, and that the bulk of the primary X-ray emission arises from regions at intermediate latitudes above and below the WD equator. To mimic such a configuration, we considered a source located at an altitude $z_s = 0.5\,R_{\rm WD}$ above the disc, which corresponds to a source located at a latitude of $30^{\circ}$ on the WD surface. In this configuration, the fluorescent line EW increases significantly to 38.9\,eV. This total EW of the line from the WD accretion disc increases to 44.2\,eV if the density at the inner disc boundary is doubled compared to the results of \citet{Ras25}\footnote{If we increase the density by a factor  of four with respect to our baseline value, the EW actually decreases to 33.3\,eV because the disc becomes optically thick over a wider range in radii.}. The EW also depends on $z_s$: when $z_s$ increases from 0.2 to 0.7\,$R_{\rm WD}$, that is when the primary X-ray source moves from an $11^{\circ}$ latitude on the WD surface to $44^{\circ}$, the total EW of the fluorescent line produced in the WD accretion disc increases from 23.7 to 41.3\,eV (for a density at the inner boundary of $3 \times 10^{16}$\,cm$^{-3}$) as a larger part of the disc is illuminated.
\begin{figure}
  \resizebox{8.5cm}{!}{\includegraphics[angle=0]{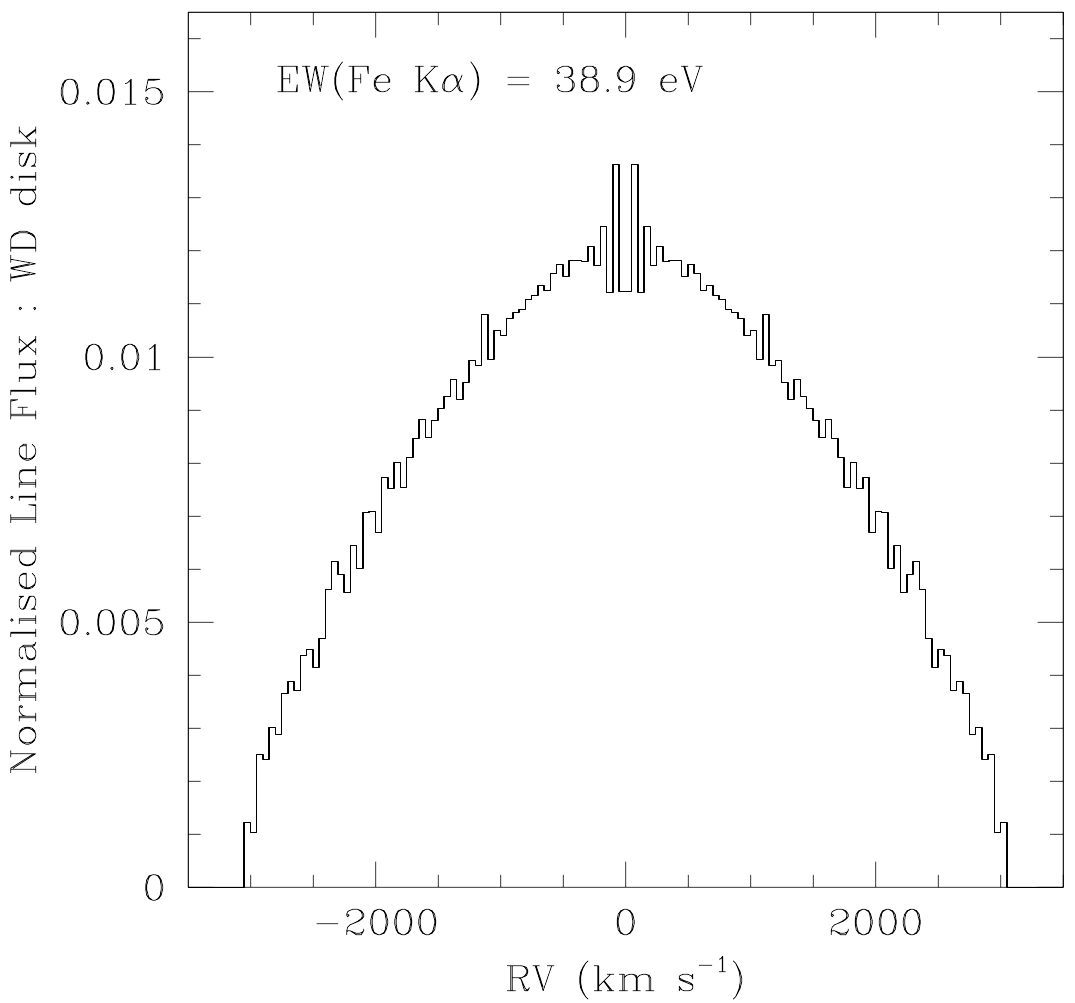}}
  \resizebox{8.5cm}{!}{\includegraphics[angle=0]{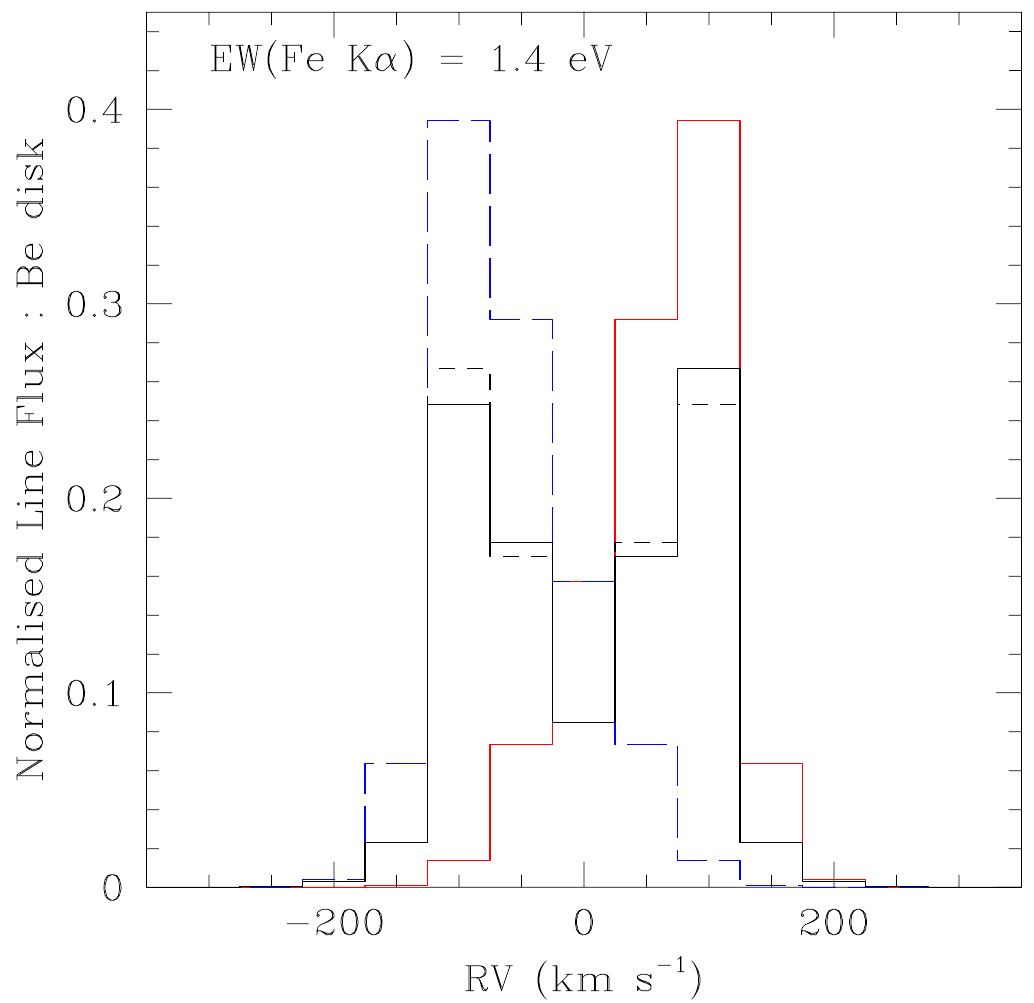}}
  \caption{Top: Normalised individual fluorescent line profile of the Fe K$\alpha$ line arising from the accretion disc around the WD in velocity bins of 50\,km\,s$^{-1}$. The primary X-ray source is located on the WD at a latitude of $30^{\circ}$ (i.e.\ $z_s = 0.5\,R_{\rm WD}$). Bottom: Normalised individual fluorescent Fe K$\alpha$ line contributed by the Be decretion disc illuminated by the WD at four orbital phases. The outer Be disc was assumed to follow the structure described by \citet{Rub25}. The black histograms stand for the conjunction phases with the WD behind (solid line) or in front (dashed line). The blue and red histograms stand for the quadrature phases with the WD moving respectively towards us and away from us.\label{profile_WDnonB}}
\end{figure}

While the situation might at first sight seem reminiscent of the fluorescence by the Be disc produced by an extended source near the Be star (Sect.\,\ref{stardisc}), Fig.\,\ref{profile_WDnonB} reveals an important difference in the shape of the line profile. Indeed, the first cosmic velocity\footnote{That is the Keplerian velocity on a circular orbit of radius equal to the stellar radius.} of the WD amounts to 4325\,km\,s$^{-1}$, which is nearly 8 times the value of the first cosmic velocity of the Be star. Hence, individual fluorescent line components formed via illumination of the WD accretion disc are expected to be considerably broader than those from the magnetic interaction scenario. Therefore, we find a total width of the individual line profiles of about 6000\,km\,s$^{-1}$ (Fig.\,\ref{profile_WDnonB}). The line profile is also quite different, displaying a roughly parabolic shape. Another difference concerns the centroid of the line. The fluorescent line from a WD accretion disc should follow the orbital motion of the WD which is expected to have a semi-amplitude near 60\,km\,s$^{-1}$, that is, 15 times larger than the RV semi-amplitude of the Be star. 

In addition to the fluorescent emission from the WD accretion disc, we must consider the contribution from the Be decretion disc. As shown in Paper I, when illuminated from the outside by a compact source, the Be decretion disc produces a contribution whose shape and centroid vary with orbital phase. In Paper I, a simple axisymmetrical decretion disc was assumed and the total EW of the ensuing fluorescent Fe K$\alpha$ component was estimated to be 1.35\,eV. In our present calculations, we account for the more complex morphology of the Be disc as described by \citet{Rub25}. The resulting individual fluorescent line profiles at four orbital phases are displayed in Fig.\,\ref{profile_WDnonB}. Compared to the results of Paper I, the profiles at conjunction phase now present a double-peaked morphology instead of a box-shaped profile. This double-peaked shape reflects the presence of the spiral arms in the outer parts of the disc. However, the EW contributed by the fluorescence from the Be disc remains at a modest value, 1.41\,eV.

\begin{figure}
  \resizebox{8.5cm}{!}{\includegraphics[angle=0]{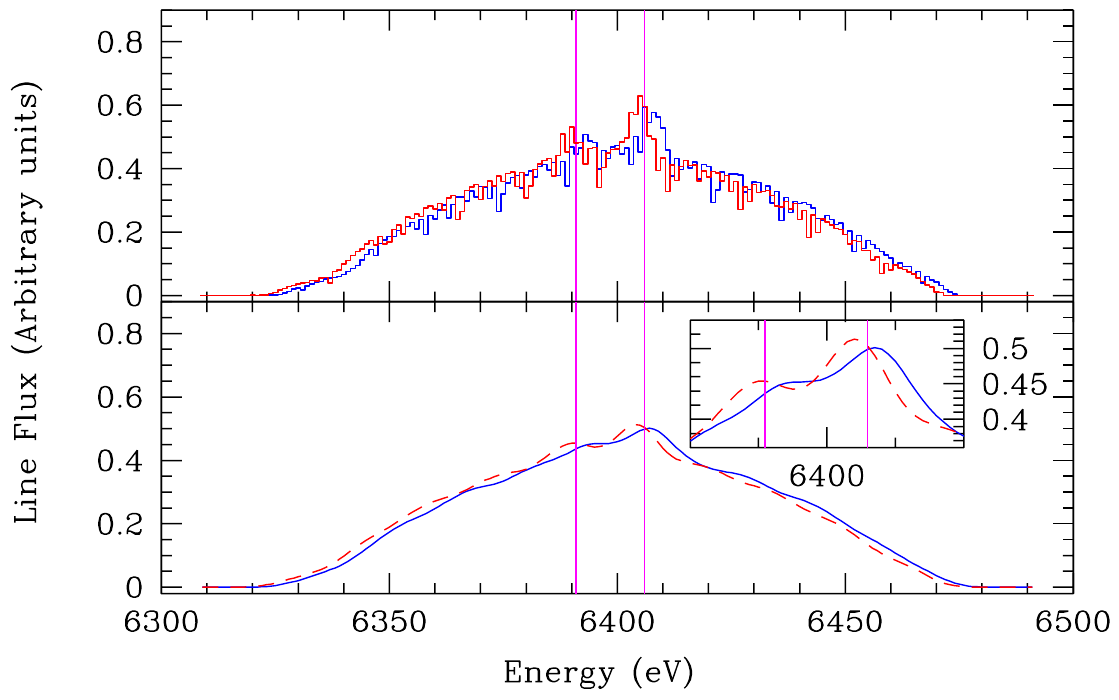}}
  \caption{Top: Global line profile (i.e.\ combining the contributions from the WD's accretion disc and the Be decretion disc and accounting for the full array of transitions) for our default non-magnetic WD model, at two opposite quadrature orbital phases. The profiles are shown as histograms with energy bins of 1\,eV. Bottom: Convolution of the global line profiles with an instrumental response function of 5\,eV FWHM. The insert displays a zoom on the energy range 6480 -- 6420\,eV, to highlight the $\sim 2.6$\,eV shift between the profiles at the extreme orbital phases. The magenta lines yield the theoretical energies of the centroids of the Fe\,{\sc v} K$\alpha$1 and K$\alpha$2 subgroups. \label{global_WDnonB}}
\end{figure}

Adding the two contributions, we conclude that an accreting non-magnetic WD scenario can actually reproduce line EWs around 30 -- 50\,eV, that is in the range of the observational values, provided that the primary X-ray source is located at intermediate latitudes on both sides of the WD equator. Figure\,\ref{global_WDnonB} displays the expected global K$\alpha$ profile at both quadrature orbital phases. Because of the considerable width of the individual profiles, the FWHM of the overall profiles observed with an instrument with a resolution of 5\,eV amounts to about 83\,eV as opposed to $\sim 23$\,eV in the magnetic interaction scenario. Finally, because the WD disc follows the WD's orbital motion, the shift of the global line profiles between the two quadrature phases is about $2.6$\,eV, a value within reach of the {\it Resolve} instrument. 

\subsection{Accretion onto a magnetic white dwarf}
In the accreting magnetic WD model, the primary X-ray emission is thought to arise in the shock-heated plasma of the post-shock accretion column \citep[PSAC, see e.g.][]{Hay14}. Immediately behind the shock, the plasma temperature follows from the pre-shock velocity, $v_{\rm pre-shock}$, via the Rankine-Hugoniot relations for strong shocks \citep[e.g.][]{Ezu99}. Assuming a ballistic flow, from the Alfv\'en radius to the standoff shock, the post-shock temperature becomes
  \begin{eqnarray}
    \label{ballistic}
    k\,T_s & = & \frac{3}{16}\,\mu\,m_{\rm H}\,v_{\rm pre-shock}^2 \nonumber \\
   &  = &\frac{3}{16}\,\mu\,m_{\rm H}\,\frac{2\,G\,M_{\rm WD}}{R_{\rm WD}}\,\left(\frac{R_{\rm WD}}{R_{\rm WD} + h_s} - \frac{R_{\rm WD}}{R_A}\right),
  \end{eqnarray}
  where $\mu$ is the mean molecular weight and $m_{\rm H}$ the mass of the hydrogen atom. Taking, for instance,  $R_A = 20\,R_{\rm WD}$ and $h_s = 0.5\,R_{\rm WD}$ results in an expected plasma temperature of 27\,keV. This is well above the observational $kT = 12.5$\,keV. As the shock-heated material settles towards the WD surface, its temperature rapidly decreases and its density increases. Hydrodynamic models of the PSAC indicate that the regions close to the WD surface, which have the highest densities and thus the highest emission measure, display significantly lower plasma temperatures \citep[see Fig.\,2 of][]{Hay14}. The temperature and density stratification of the PSAC yields an X-ray spectral energy distribution \citep{Tsu18,Tsu23} that mimics that of conventional optically thin thermal plasma models of significantly lower $kT$ \citep{Rau22}. For the simulations in this section, we kept the plasma temperature at $kT = 12.5$\,keV, that is the mean value obtained from broadband X-ray spectral fitting of $\gamma$~Cas \citep{Rau22}. This value should thus not be seen as the immediate post-shock temperature, but rather as an emission-measure-weighted average along the PSAC. In the same way, the $h_s$ parameter in our simulations should not be taken as the actual height of the standoff shock, but represents instead the height where most of the primary X-ray emission arises.

In this scenario, we must consider four reservoirs of fluorescent material. These are the WD photosphere, the truncated accretion disc, the pre-shock accretion flow and the circumstellar material of the Be star.  

We first consider the contribution of the WD atmosphere. For this purpose, we used version 208 of the TLUSTY nLTE model atmosphere code \citep{Hub17a,Hub17b,Hub17c,Hub21} to compute a WD model atmosphere with $T_{\rm eff} = 60\,000$\,K and $\log{g} = 8.5$. We considered three different situations. First, we assumed a hot DA WD with a nearly pure H atmosphere containing only traces of C ($0.01 \times$ solar) and Fe ($0.4 \times$ solar). As a second step, we considered a DO WD with a He abundance a hundred times that of hydrogen and the same traces of C and Fe \citep{Nou07}. For each case, we first built a pure H (or pure He + H) LTE model atmosphere. This model was then used as the starting point of an iterative process to compute a nLTE model. Finally, C and Fe were incorporated and the model was iterated to obtain a full nLTE solution \citep{Hub17c}. In addition, we considered the moderately cool pure-H DA WD model ($T_{\rm eff} = 10\,000$\,K, $\log{g} = 8.0$) of \citet{Hub17c}. 

For the two hot WD model atmospheres, the optical depth for photoelectric absorption of X-rays with energies above the Fe K edge reaches unity at higher Rosseland optical depths than in the case of the Be star (see Fig.\,\ref{Rosseland}). In the case of the DA WD, this occurs at $\tau_{\rm Ross} \simeq 25$, whereas it happens at $\tau_{\rm Ross} \simeq 56$ for the DO WD model. This difference with respect to the Be star is a consequence of the higher temperature and the low metal contents of the WD models. Both effects reduce the Rosseland opacity. According to our TLUSTY models, the dominant Fe ions at these optical depths are Fe\,{\sc v} for the DA WD and Fe\,{\sc vi} for the DO WD. Some studies of DA WDs indicate that metals might not be homogeneously distributed in the WD atmosphere, but could instead follow a stratification with nearly pure H or He atmospheres at the top \citep[e.g.][]{Rauch13}. Whilst our TLUSTY models were computed with rather low iron abundances and assuming a chemically homogeneous atmosphere without stratification, in the computation of the source function due to fluorescence, we assume a solar abundance. This assumption was introduced to account for stratification resulting in a higher Fe abundance at the $\tau_{\rm Ross}$ where fluorescence occurs, and to ease comparison with other calculations in this paper which all assume solar Fe abundances. 

\begin{figure}
  \resizebox{8.5cm}{!}{\includegraphics[angle=0]{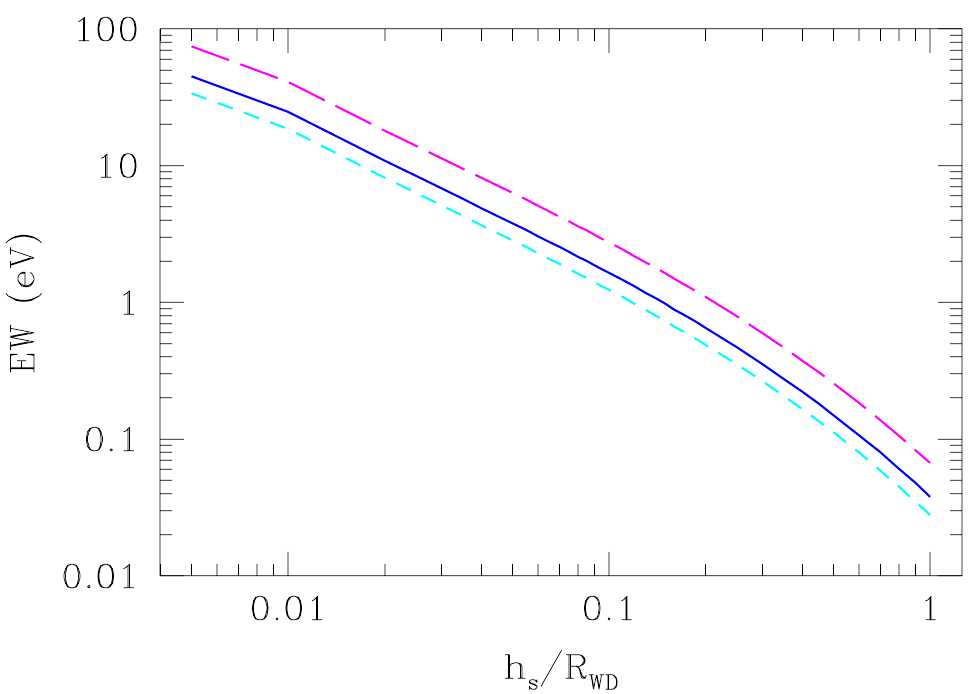}}
  \resizebox{8.5cm}{!}{\includegraphics[angle=0]{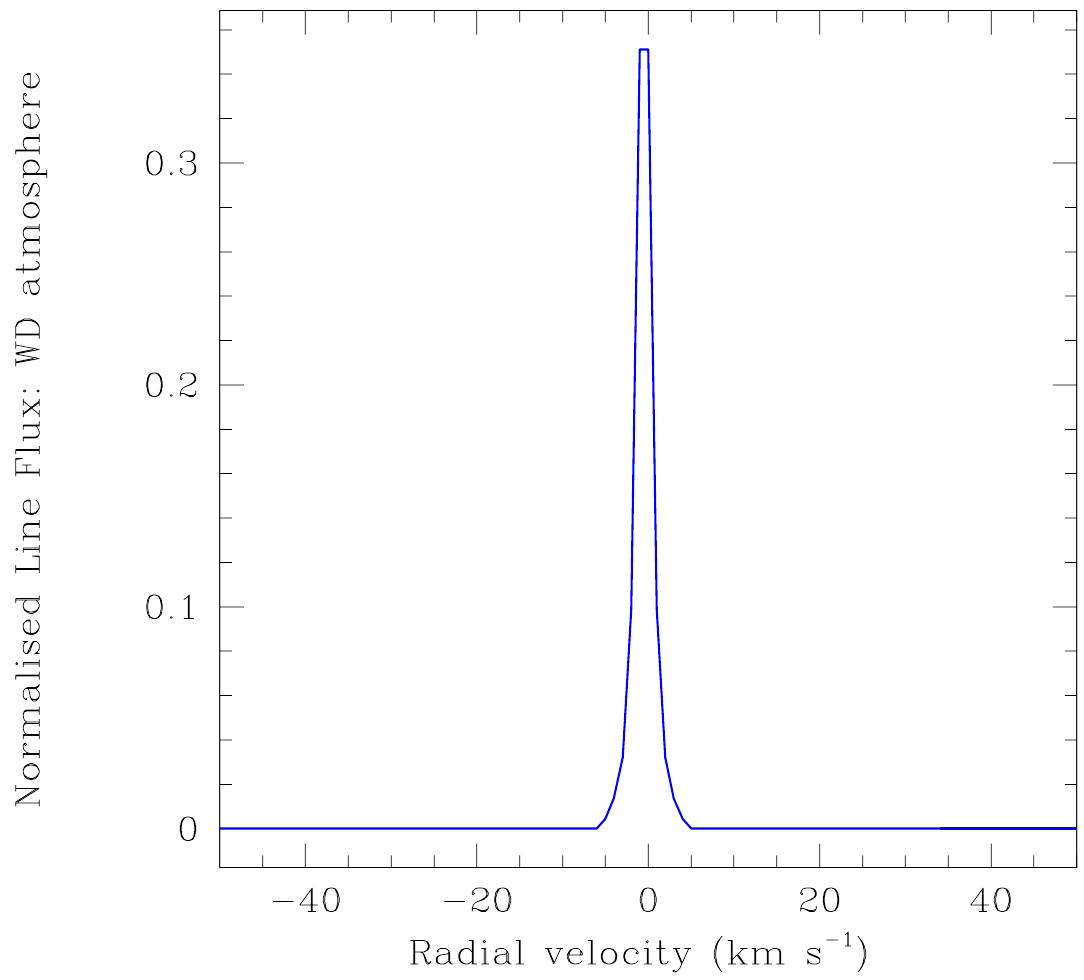}}
  \caption{Top: EW of the Fe K$\alpha$ fluorescent line due to reflection on the WD surface as a function of the height of the accretion shock above the WD surface. The short-dashed cyan, solid blue, and long-dashed magenta lines correspond  to the hot DA, the DO, and the moderately cool DA WD models, respectively. In all cases, an inclination of $45^{\circ}$ was assumed. Bottom: Synthetic profile for an individual component of the Fe K$\alpha$ fluorescent line produced in the hot DA WD atmosphere for $v_{\rm rot}^{\rm WD} = 25$\,km\,s$^{-1}$, $h_s = 0.04\,R_{\rm WD}$, $i=45^{\circ}$, and accretion taking place at the magnetic pole assumed to coincide with the rotational pole.\label{EWWD}}
\end{figure}

We then used these WD model atmospheres in combination with our Feautrier method (Sect.\,\ref{SSatmo} and Appendix\,\ref{Feautrier}) to compute the WD's photospheric contribution to the fluorescent line. We assume accretion near the WD's magnetic poles which we consider, for simplicity, to coincide with the rotational poles. We further assume that the rotation and orbital axes are aligned and are seen under an angle $i = 45^{\circ}$ by the observer. We computed the EWs for different values of $h_s$ and considering all three WD atmosphere models (see top panel of Fig.\,\ref{EWWD}). The $h_s$ parameter was varied between 0.005\,$R_{\rm WD}$ and 1.0\,$R_{\rm WD}$ in steps of 0.005\,$R_{\rm WD}$ for $h_s/R_{\rm WD}$ between 0.005 and 0.1, 0.01\,$R_{\rm WD}$ for $h_s/R_{\rm WD}$ between 0.1 and 0.2, 0.05\,$R_{\rm WD}$ for $h_s/R_{\rm WD}$ between 0.2 and 0.5, and 0.1\,$R_{\rm WD}$ for $h_s/R_{\rm WD}$ between 0.5 and 1.

For a primary X-ray source located at $h_s = 0.04\,R_{\rm WD}$ above the WD's pole, we obtain EWs of 3.68\,eV and 4.90\,eV respectively for the DA and DO WD models. This EW increases to 8.15\,eV for the moderately cool DA WD. If we assume instead an extremely small $h_s = 0.001\,R_{\rm WD}$, then the hot and moderately cool DA WD atmospheres produce EWs of respectively 45.9 and 101.5\,eV. These comparisons underline the strong dependence of the line strength on the unknown properties of the putative WD and of its accretion flow. 

Finally, as one could expect given the WD's rotational velocity, which is much lower than that of the Be star, the synthetic profiles of the individual components are extremely narrow (see Fig.\,\ref{EWWD}). This conclusion is independent of the assumptions made for the WD atmospheric model and depends only on the adopted projected rotational velocity. 

We can compare our results with those of \citet{vanTee96} who considered the formation of the fluorescent K$\alpha$ line via isotropical illumination of the WD atmosphere. In the absence of a strong temperature inversion, which occurs above a threshold value of the illuminating flux (see below), \citet{vanTee96} predict an EW by reflection on the WD of about 57\,eV for a WD with an effective temperature near 50\,kK and considering a primary bremsstrahlung radiation with $kT = 10$\,keV. Our EW values are much lower than those reported by \cite{vanTee96}. The reasons for this difference lie in their assumption of a fluorescent medium occupying a $2\,\pi$ solid angle, as seen from the primary source, and undergoing isotropic illumination, that is with all fluorescent elements located at the same distance from the primary source. However, the actual geometry is that of a point source illuminating a sphere from the outside. As a result, only a limited cap with an extension in polar angle from 0 to $\arccos{\frac{R_{\rm WD}}{R_{\rm WD} + h_s}}$ of the WD surface will be illuminated. For $h_s = 0.04\,R_{\rm WD}$, this cap intercepts a solid angle of $0.725 \times 2\,\pi$ as seen from the primary source. Moreover, the majority of the illuminated parcels are farther away from the primary source than the parcel at the nadir point of the primary source and will thus receive a diluted flux. For $h_s = 0.04\,R_{\rm WD}$, the geometrical dilution produces a reduction of the EW by a factor 7. Finally, the EW obviously also depends on the viewing angle. Adopting $i = 90^{\circ}$ instead of $45^{\circ}$ yields a 40\% increase of the EW. Taken together, these effects can thus easily account for the lower EWs in our calculations compared to those of \citet{vanTee96}.

Additional uncertainties affect the level of the WD's photospheric contribution. For instance, \citet{vanTee96} drew attention to the fact that the optically thin layers at the top of an irradiated WD atmosphere can be subject to a temperature inversion if the primary flux exceeds $10^{16}$\,erg\,cm$^{-2}$\,s$^{-1}$. This leads to the occurrence of a range of ionisation stages resulting in fluorescent K$\alpha$ line emission for ions up to Fe\,{\sc xxii}. Because the energy of the lines increases for the highest ionisation stages, the energy domain between 6.4 and 6.6\,keV gets populated with weaker fluorescent emissions. Hence the overall EW and the line width increase, and the centre of the whole fluorescent complex shifts towards higher energies. The maximum observed luminosities of $\gamma$~Cas stars \citep{Smi16,Naz18} yield an incident flux exceeding the threshold for temperature inversion for cases where $h_s < 0.16\,R_{\rm WD}$. Temperature inversion is currently not accounted for in our TLUSTY models which consider the X-rays as a perturbation of the atmosphere that does not alter its temperature and density structure.\\ 

\begin{figure}
  \resizebox{8.5cm}{!}{\includegraphics[angle=0]{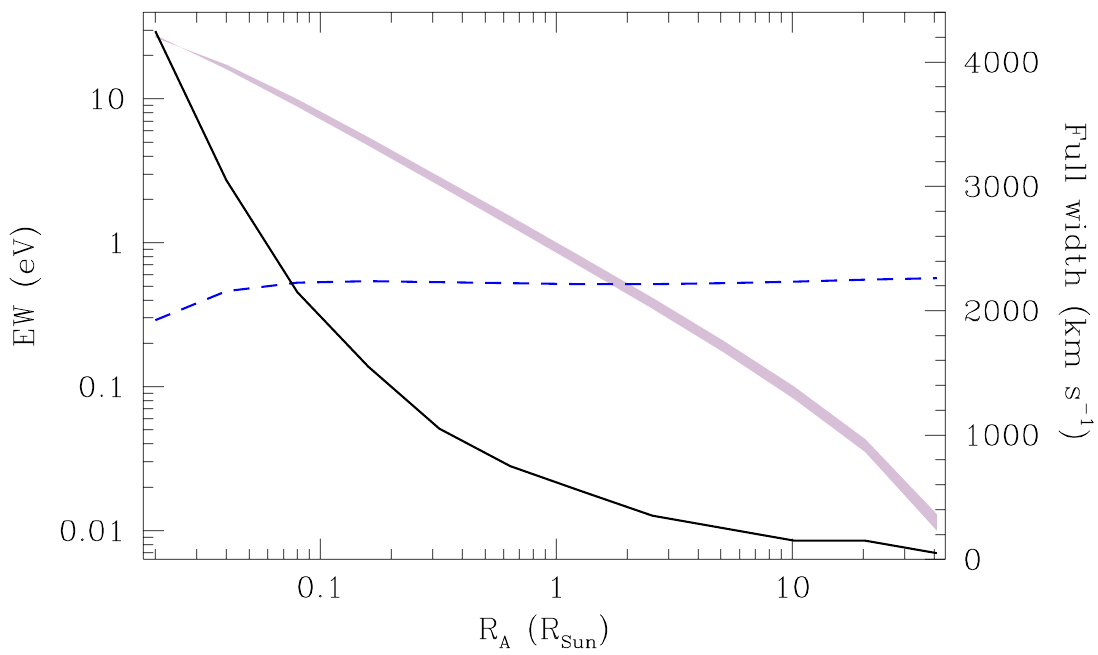}}
  \caption{Dependence of the EW and line width of individual fluorescent components from the truncated accretion disc and accretion flow on the WD Alfv\'en radius, $R_A$. The grey-shaded area, with the corresponding vertical axis on the left, yields the total EW contributed by the accretion disc for values of $h_s$ ranging between 0.005\,$R_{\rm WD}$ and 1\,$R_{\rm WD}$. The dashed blue curve indicates the EW contributed by the accretion flow. The solid black line yields the full width of the individual lines formed in the truncated accretion disc with the corresponding vertical axis on the right.\label{EW_discWD_Bfield}}
\end{figure}

The second most important contribution to the fluorescent line for the accreting magnetic WD scenario comes from the truncated accretion disc. Obviously, the WD Alfv\'en radius plays a key role here. Figure\,\ref{EW_discWD_Bfield} illustrates the EW and the line width for individual fluorescent line components as a function of $R_A$. The EW of this contribution is nearly insensitive to $h_s$, but rapidly decreases from values near 27.6\,eV, for $R_A = 2\,R_{\rm WD} = 0.02$\,R$_{\odot}$, to 8.8\,eV, for $R_A = 8\,R_{\rm WD} = 0.08$\,R$_{\odot}$, and $< 1$\,eV, for $R_A \geq 80\,R_{\rm WD} = 0.8$\,R$_{\odot}$. The full width of the line is independent of $h_s$ and also displays a steep decrease with $R_A$. This is expected as the truncation of the inner disc removes the material with the fastest orbital motion around the WD.\\    

Finally, we address the contribution to the fluorescent line arising in the material of the magnetically channelled accretion flow. We assume here that the material flows along the magnetic field lines and the magnetic field can be described as dipolar. The field lines connect the inner boundary of the accretion disc, truncated at the Alfv\'en radius, with the magnetic pole of the WD. Assuming the relevant field line to be located in the $(x,z)$ plane, we have $x = R_A\,\sin^3{\theta}$, $z = R_A\,\sin^2{\theta}\,\cos{\theta}$, and $r = \sqrt{x^2+z^2} = R_A\,\sin^2{\theta}$. Following \citet{Hay21}, we assume that the material moves at the free fall velocity, $v = \sqrt{\frac{2\,G\,M_{\rm WD}}{r}}$, and that the accretion flow has a cross-section $S = S_0\,\left(\frac{r}{R_{\rm WD}}\right)^3$. Here $S_0$ is the cross-section at the surface of the WD which is directly related to the accretion rate $\alpha_{\rm acc}$ by the expression
\begin{equation}
  \dot{M}_{\rm acc} = \rho\,v\,S = \alpha_{\rm acc}\,S_0
.\end{equation}
\citet{Tsu18} applied the {\tt acrad} model of \citet{Hay18} for a magnetic accreting WD to the {\it NuSTAR} and {\it Suzaku} observations of $\gamma$~Cas. In this model, the X-ray spectrum is computed as a function of WD mass by solving the plasma fluid equation along the magnetic field line and dealing with reprocessing of the X-rays that reflect off the WD surface\footnote{The WD surface is approximated as a solid in this model.}. In this way, \citet{Tsu18} inferred $\dot{M}_{\rm acc} = 1.3 \times 10^{-10}$\,M$_{\odot}$\,yr$^{-1}$ and $\log{\alpha_{\rm acc}} = -0.61$, where $\alpha_{\rm acc}$ is expressed in units g\,cm$^{-2}$\,s$^{-1}$. This accretion rate is 6.5 times higher than the value found in the SPH calculations of \citet{Ras25} and corresponds to a hydrogen number density of $1.8 \times 10^{14}$\,cm$^{-3}$ at the WD surface. To infer an optimistic value of the contribution of the accretion flow, we thus adopt the \citet{Tsu18} values. The number density decreases with distance from the WD as $\left(\frac{R_{\rm WD}}{r}\right)^{5/2}$ \citep{Hay21}, whilst the volume occupied by the flow increases as $r^4$. We assume that accretion flows take place in both magnetic hemispheres. 

\begin{figure}
  \resizebox{8.5cm}{!}{\includegraphics[angle=0]{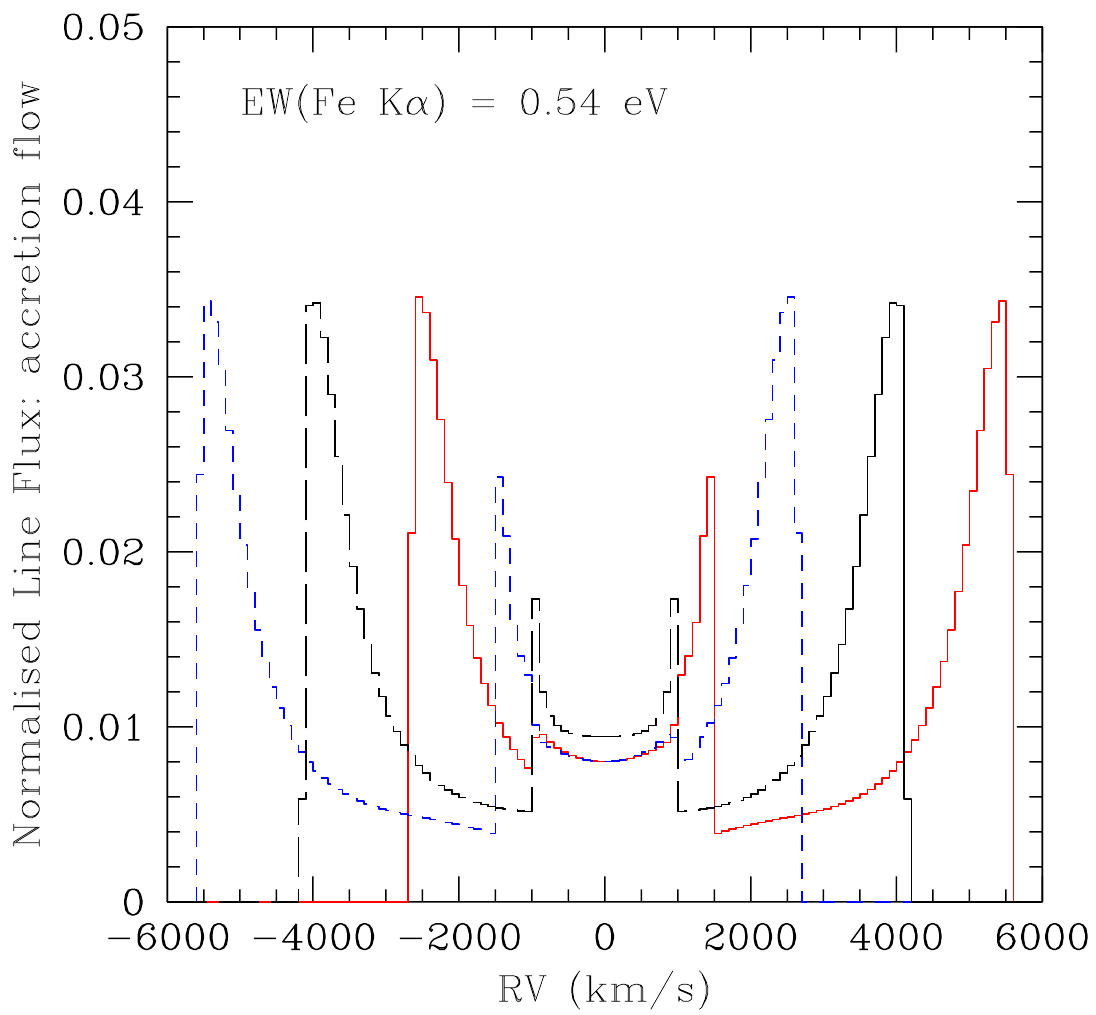}}
  \caption{Illustration of the expected individual line profiles for Fe K$\alpha$ fluorescent lines arising from illumination of the material in the accretion flow. The Alfv\'en radius, $R_A$, was taken to be $20\,R_* = 0.2$\,R$_{\odot}$. The different lines correspond to different azimuthal rotation angles of the magnetosphere. The observer is located in the direction corresponding to the positive $(x,z)$ plane and seeing the system under an inclination of $i = 45^{\circ}$. The WD is located at the origin of the axes. The line colours correspond to orientations of the accretion flow located inside the positive $(x,z)$ plane (continuous red line), the $(y,z)$ plane (dashed black line), and the negative $(x,z)$ plane (short dashed blue line).\label{curtain}}
\end{figure}

Because of the high escape velocity of the WD, and because we assume that flows occur along field lines in both hemispheres, the contribution to the fluorescent line profile displays a broad profile with a double-peaked morphology (see Fig.\,\ref{curtain}). The peaks are quite narrow and their location depends on the aspect angle of the field lines with respect to the line of sight. The profiles are thus expected to change with the rotation phase of the WD's magnetosphere (see Fig.\,\ref{curtain}). The EW of this contribution remains quite modest, however. For small Alfv\'en radii, the EW first increases with $R_A$ but quickly reaches a nearly constant value around 0.54\,eV for $R_A \geq 0.08$\,R$_{\odot}$.\\

Finally, fluorescence from the illumination of the Be disc by the primary X-ray source located at the WD position is also included in our model (see Paper I). As for the non-magnetic WD case, this component yields rather low contribution of 1.4\,eV to the EW.

\begin{figure}
  \resizebox{8.5cm}{!}{\includegraphics[angle=0]{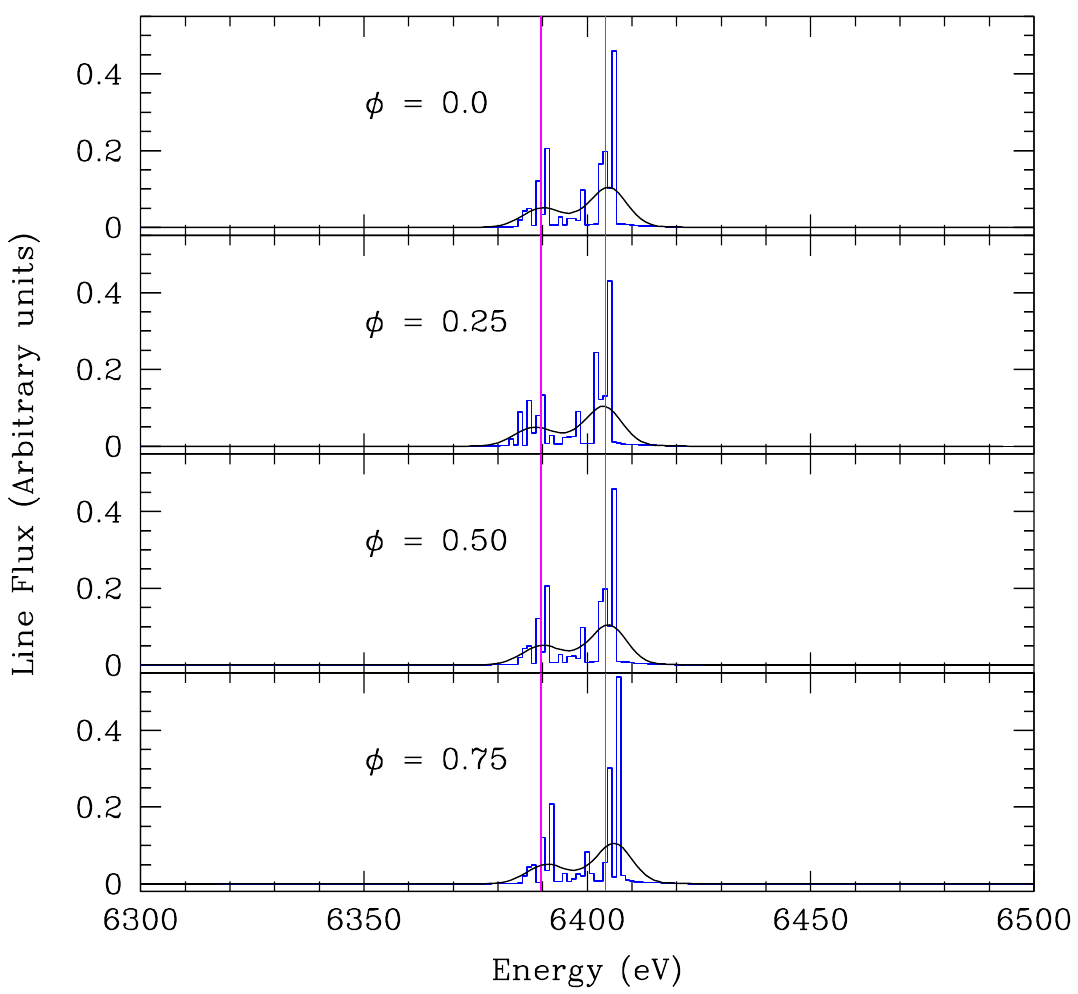}}
  \caption{Global Fe K$\alpha$ lines for the case of an accreting magnetic WD assuming an EW of 50\,eV for the component arising via reflection on the WD. We assume an Alfv\'en radius of $20\,R_{\rm WD} = 0.2$\,R$_{\odot}$. The blue histograms show the profiles with 1\,eV energy bins, whereas the black lines show the results of a convolution with an instrumental response function of 5\,eV FWHM. The magenta lines indicate the theoretical energies of the centroids of the Fe\,{\sc vi} K$\alpha$1 and K$\alpha$2 subgroups (i.e.\ assuming Fe {\sc v} target ions). We note the changes in energy of the line peaks, which amount to 2.6\,eV between opposite quadrature phases.\label{globalWDwithBstrongrefl}}
\end{figure}

Our ignorance of the properties of the putative WD and of the location of the accretion shock prevents us from computing all the ingredients of the magnetic WD model self-consistently. Nonetheless, we again combine all these ingredients to build a global line profile. For this purpose, we assume that the photospheric contribution has an EW of 50\,eV, close to the observed EWs, and take a medium value of $R_A = 20\,R_{\rm WD}$. We then proceed as for the other scenarios to build the global Fe K$\alpha$ line profile. The result is shown in Fig.\,\ref{globalWDwithBstrongrefl}. Whilst this specific model is not necessarily representative of the companion of $\gamma$~Cas or any of its siblings, our calculations in this section nonetheless provide some important insight. First of all, the magnetic WD scenario can produce fluorescent lines with the strengths observed in $\gamma$~Cas stars. The actual strength depends strongly on the properties of the WD atmosphere and to a lesser extent on its magnetic field strength. Our calculations suggest that cooler WDs are more efficient in producing fluorescent Fe K$\alpha$ lines. However, given the expected youth of the WD in $\gamma$~Cas which presumably formed as the outcome of a mass-exchange episode, the hot DA or DO WD models are probably more appropriate to describe the WDs associated with early-type Be stars. Second, synthetic line profiles are considerably narrower than those expected for a non-magnetic WD. The overall width is very similar to what we found for the magnetic star-disc interaction scenario. Finally, the centroids of the lines change with orbital phase as they are expected to follow the orbital motion of the WD, leading to a shift between quadrature phases of 2.6\,eV which is measurable with {\it Resolve} and {\it X-IFU}. 

\section{Conclusions \label{conclusion}}
In this article, we investigated the fluorescent Fe K$\alpha$ line profiles as they will be seen in high-resolution X-ray spectra of $\gamma$~Cas stars according to the most-plausible scenarios: magnetic star-disc interaction or accretion by a WD companion. Our simulations indicate that the observable line profiles provide strong constraints on the origin of the hard primary X-ray emission. The most important constraints are summarised in Table\,\ref{Tabfinal}.

\begin{table}
  \caption{Summary of expected fluorescent Fe K$\alpha$ line properties\label{Tabfinal}}
  \begin{center}
    \begin{tabular}{c | c  | c}
      \hline
    Scenario & Line orbital motion & Line width \\
    \hline
    Star-disc interaction & follows Be star & narrow \\
    Non-magnetic WD & follows WD & broad \\
    Magnetic WD & follows WD & narrow \\
    \hline
  \end{tabular}
  \end{center}
\end{table}

For a non-magnetic accreting WD, the fluorescence mostly occurs in the WD accretion disc, close to the WD itself. Hence, the predicted global Fe K$\alpha$ complex is very broad with a full width near 140\,eV. In strong contrast with this situation, for the magnetic star-disc interaction and the magnetic accreting white dwarf cases, the full width of the Fe K$\alpha$ complex is expected to be $\sim 40$\,eV. This offers the first powerful diagnostic to distinguish the non-magnetic WD case from the other two scenarios. 

Despite the very similar shapes of the overall line profiles, the magnetic star-disc interaction scheme and the magnetic WD scenario can also be distinguished thanks to the fluorescent line. Indeed, in the magnetic star-disc interaction, the inner decretion disc of the Be star provides the main reservoir of fluorescent material, which implies that the centroids of the Fe K$\alpha$ transitions follow the orbital motion of the Be star. On the contrary, in the magnetic accreting WD configuration, the bulk of the fluorescence comes from the WD photosphere with some additional contributions from the WD accretion disc and the magnetically channelled accretion flow. In this case, the centroid of the line complex thus follows the motion of the WD, which is in anti-phase with the motion of the Be star and has a significantly larger amplitude (around 2.6\,eV for $\gamma$~Cas). 

Comparing upcoming {\it XRISM/Resolve} observations of $\gamma$~Cas at some well-chosen orbital phases (orbital quadratures) with our synthetic line profiles will provide a major step forwards in the understanding of the mysterious X-ray emission of $\gamma$~Cas stars. Provided the amplitudes of the expected RV excursions are larger than the instrumental precision and the observational data are of sufficient quality to determine precise line widths and centroids, {\it XRISM} observations at several well-selected orbital phases should enable us to differentiate between the competing scenarios. Whilst {\it XRISM} observations will be restricted to the X-ray-brightest objects of the $\gamma$~Cas category, {\it Athena/X-IFU} should be able to collect such data for a larger number of $\gamma$~Cas stars, with different stellar and binary properties. This will allow us to sample the $\gamma$~Cas population at large and enable a better understanding of the evolutionary status of these systems.  

\begin{acknowledgements}
  This study was partially supported by the BELgian federal Science Policy Office (BELSPO) and the European Space Agency through the PRODEX HERMeS grant. ADS and CDS were used during this research. My deepest gratitude goes to Drs Ya\"el Naz\'e, Myron Smith, Christian Motch, and Masahiro Tsujimoto for discussion, and to the referee for remarks that helped me improve this work.
\end{acknowledgements}

\bibliographystyle{aa}
\bibliography{mybiblio}
  \begin{appendix}
\section{Fluorescence in the stellar atmosphere \label{Feautrier}}
In a plane-parallel atmosphere, we have
\begin{equation}
  \mu\,\frac{\partial I_E(z,\mu,E)}{\partial\,z} = \rho(z)\,\eta(z,E) - \rho(z)\,\chi(z,E)\,I_E(z,\mu,E), \label{eqB1}
  \end{equation}
with $I_E(z,\mu,E)$ the specific intensity per unit photon energy at energy $E$ at coordinate $z$ and propagating in the direction $\mu$, $\rho(z)$ is the gas density, $\eta(z,E)$ and $\chi(z,E)$ are respectively the emissivity and opacity at energy $E$ and depth $z$.

Using the Rosseland optical depth $d\tau_{\rm Ross} = -\kappa_{\rm Ross}\,\rho\,dz$, Eq.\,(\ref{eqB1}) becomes
\begin{equation}
  \mu\,\varpi\,\frac{\partial I_E(\tau_{\rm Ross},\mu,E)}{\partial\,\tau_{\rm Ross}} = I_E(\tau_{\rm Ross},\mu,E) - S(\tau_{\rm Ross},E),
  \label{Feaut1}
  \end{equation}
where $\varpi = \frac{\kappa_{\rm Ross}(\tau_{\rm Ross})}{\chi(\tau_{\rm Ross},E)}$, and $S(\tau_{\rm Ross},E)$ is the source function. Restricting $\mu$ to positive values ($0 \leq \mu \leq 1$), we can split $I_E$ into two regimes (outwards-propagating and inwards-propagating):
\begin{eqnarray}
  I^+_E(\tau_{\rm Ross},\mu,E) & = & I_E(\tau_{\rm Ross},\mu,E) \\
  I^-_E(\tau_{\rm Ross},\mu,E) & = & I_E(\tau_{\rm Ross}, -\mu,E). 
\end{eqnarray}
We then define
\begin{eqnarray}
  u & = & \frac{1}{2}\,\left(I^+_E(\tau_{\rm Ross},\mu,E) + I^-_E(\tau_{\rm Ross},\mu,E)\right),\\
  v & = & \frac{1}{2}\,\left(I^+_E(\tau_{\rm Ross},\mu,E) - I^-_E(\tau_{\rm Ross},\mu,E)\right). 
\end{eqnarray}
Hence, Eq.\,(\ref{Feaut1}) leads to
\begin{eqnarray}
  \varpi\,\mu\frac{\partial\,u}{\partial\tau_{\rm Ross}} & = & v,\\
  \varpi\,\mu\frac{\partial\,v}{\partial\tau_{\rm Ross}} & = & u - S_E(\tau_{\rm Ross},E)
,\end{eqnarray}
which finally yields
\begin{equation}
  \varpi^2\,\mu^2\,\frac{\partial^2\,u}{\partial\tau_{\rm Ross}^2} + \varpi\,\mu^2\,\frac{\partial\,\varpi}{\partial\tau_{\rm Ross}}\,\frac{\partial\,u}{\partial\tau_{\rm Ross}} = u - S_E(\tau_{\rm Ross},E). \label{eqB9}
\end{equation}
Equation\,(\ref{eqB9}) is solved with the boundary conditions
\begin{eqnarray}
  z = 0 & \Rightarrow & u_E(0,\mu) - \varpi\,\mu\,\frac{\partial\,u_E(0,\mu)}{\partial\tau_{\rm Ross}} = I_{X,E}(\mu), \\
  z \rightarrow -\infty & \Rightarrow & u_E(\infty,\mu) + \varpi\,\mu\,\frac{\partial\,u_E(\infty,\mu)}{\partial\tau_{\rm Ross}} = 0,
\end{eqnarray}
where $I_{X,E}$ is the illuminating specific intensity per unit photon energy which is non-zero only for the value of $\mu$ corresponding to the angle of illumination by the external X-ray source. We follow \citet{Gar13} in expressing this quantity as
\begin{equation}
  I_{X,E} = \frac{2\,F(E)}{\mu_{\rm in}}\,\delta(\mu-\mu_{\rm in}),
\end{equation}
where $F(E)$ is the flux at energy $E$ illuminating the photosphere. 

The source function has two contributions:\\
$\bullet$ electron scattering which we treat as in \citet{Gar10} with
  \begin{equation}
    S_{\rm scat} = \frac{\alpha_{KN}\,n_e}{\chi(\tau_{\rm Ross},E)}\,\int_0^{\infty} \frac{J_{E'}}{\sigma\,\sqrt{\pi}}\,\exp{\left(-\frac{(E-E_C')^2}{\sigma^2}\right)}\,dE'
  ,\end{equation}
  where $\alpha_{KN}$ is the Klein-Nishina cross-section, $E_C' = E'\left(1 + \frac{4\,kT - E'}{m_e\,c^2}\right)$, $\sigma^2 = E'^2\,\left(\frac{2\,kT}{m_e\,c^2} + \frac{2}{5}\,\left(\frac{E'}{m_e\,c^2}\right)^2\right)$, and $J_{E'} = \int_0^1 u_{E'}(\mu)\,d\mu$ is the mean intensity per unit photon energy at an energy $E'$.\\
$\bullet$ fluorescent emission which we express as
  \begin{equation}
    S_{\rm fluor} = \frac{z_{\rm Fe}\,x_{\rm ion}\,\omega_{\rm ion+1}\,n_{\rm H}}{\chi\,\rho}\,\int_{\rm E_{\rm thres}}^{\infty} \sigma_{\rm K-shell}(E')\,\frac{J_{E'}}{E'}\,dE'
  ,\end{equation}
  where $\sigma_{\rm K-shell}(E')$ is the K-shell photo-ionisation cross-section computed according to \citet{Ver95}, $E_{\rm thres}$ is the threshold energy for K-shell ionisation, $z_{\rm Fe}$ is the abundance of iron with respect to hydrogen \citep{Asp09}, $x_{\rm ion}$ is the relative abundance of a specific Fe ion, and $\omega_{\rm ion+1}$ the associated K$\alpha$ fluorescent yield \citep{Kal04}.
  
The above set of equations is solved in an iterative way. The iterations are stopped when the $\chi^2$ evaluated from the relative differences of $J_E$ in consecutive iterations drop below $10^{-5}$. Convergence depends on the value of $\mu_{\rm in}$, but is usually achieved within less than ten iterations. The outwards propagating specific intensity is directly proportional to the intensity of the incoming X-ray emission and is not sensitive to the initial approximation of $J_{E'}$ (either taken to be zero or equal to the intensity of the incoming X-rays multiplied by $\exp{(-\tau_{\rm Ross})}$).
\end{appendix}
\end{document}